\documentclass{article}

\PassOptionsToPackage{numbers,sort&compress}{natbib}

\usepackage[utf8]{inputenc}
\usepackage[T1]{fontenc}
\usepackage{hyperref}
\usepackage{url}
\usepackage{booktabs}
\usepackage{amsfonts}
\usepackage{nicefrac}
\usepackage{microtype}
\usepackage[table]{xcolor}
\usepackage{graphicx}
\usepackage{subcaption}
\usepackage{caption}
\usepackage{hyphenat}
\usepackage{setspace}
\usepackage{amsmath}
\usepackage{amssymb}
\usepackage{mathtools}
\usepackage{amsthm}
\usepackage{longtable}
\usepackage{multirow}
\usepackage{float}
\usepackage{array}
\usepackage{makecell}
\usepackage{tabularx}
\usepackage{enumitem}
\usepackage{threeparttable}
\usepackage{titletoc}

\usepackage[preprint]{my_style}

\usepackage{tcolorbox}
\tcbuselibrary{skins, breakable, listings}

\definecolor{brandblue}{RGB}{57,95,207}
\definecolor{linkblue}{HTML}{0064E0}
\definecolor{textgray}{HTML}{1C2B33}
\definecolor{boxbg}{HTML}{F1F4F7}
\definecolor{slate}{HTML}{E2E8F0}

\hypersetup{
  colorlinks=true,
  linkcolor=brandblue,
  citecolor=brandblue,
  urlcolor=linkblue
}

\newcommand{\method}{ComAct}
\newcommand{\bench}{ComCADBench}
\newcommand{\actor}{ComActor}

\newcommand{\modelicon}[1]{%
  \raisebox{-0.02cm}{%
    \includegraphics[height=0.20cm]{sections/images/#1}%
  }%
}

\newcommand{\paperTitle}{ComAct: Reframing Professional Software Manipulation via COM-as-Action Paradigm}

\newcommand{\paperAuthors}{%
  {\small\sffamily\bfseries
    Jiaxin Ai$^{1,2,3}$,
    Tao Hu$^{4,2}$,
    Xuemeng Yang$^{2}$,
    Shu Zou$^{5,2}$,
    Hairong Zhang$^{6,2}$,\\
    Daocheng Fu$^{6,2}$,
    Yu Yang$^{2}$,
    Hongbin Zhou$^{2}$,
    Nianchen Deng$^{2}$,
    Pinlong Cai$^{2}$,\\
    Zhongyuan Wang$^{1}$,
    Botian Shi$^{2,3}$,
    Kaipeng Zhang$^{7,3}$,
    Licheng Wen$^{2,3,\dagger}$%
  }%
}

\newcommand{\paperAffiliations}{%
  {\footnotesize
    $^1$ Wuhan University \quad
    $^2$ Shanghai Artificial Intelligence Laboratory \\
    $^3$ Shanghai Innovation Institute\quad
    $^4$ University of Science and Technology of China \\
    $^5$ The Australian National University \quad
    $^6$ Fudan University \quad
    $^7$ Alaya Studio%
  }%
}

\newcommand{\paperNotes}{%
  {\footnotesize $^\dagger$ Corresponding author.}%
}
\newif\ifshowAuthorMeta
\showAuthorMetatrue

\newcommand{\projectURL}{https://knowledgexlab.github.io/ComAct/}
\newcommand{\githubURL}{https://huggingface.co/KnowledgeXLab/ComActor-9B}
\newif\ifshowProjectLink
\newif\ifshowGithubLink
\showProjectLinkfalse
\showGithubLinkfalse
\newcommand{\publishDate}{\today}

\newcommand{\renderOptionalText}[1]{%
  \if\relax\detokenize{#1}\relax
  \else
  #1\par
  \fi
}

\newcommand{\renderFrontBox}{%
  \tcbset{
    enhanced, frame hidden,
    colback=boxbg,
    left=0.5cm, right=0.5cm, top=0.5cm, bottom=0.5cm,
    arc=16pt,
    before skip=0pt,
    grow to left by=1.5pt, grow to right by=1.5pt,
    overlay={
      \node[anchor=north east, at=(frame.north east), xshift=-2.3cm, yshift=-0.5cm]
      {\includegraphics[height=1cm]{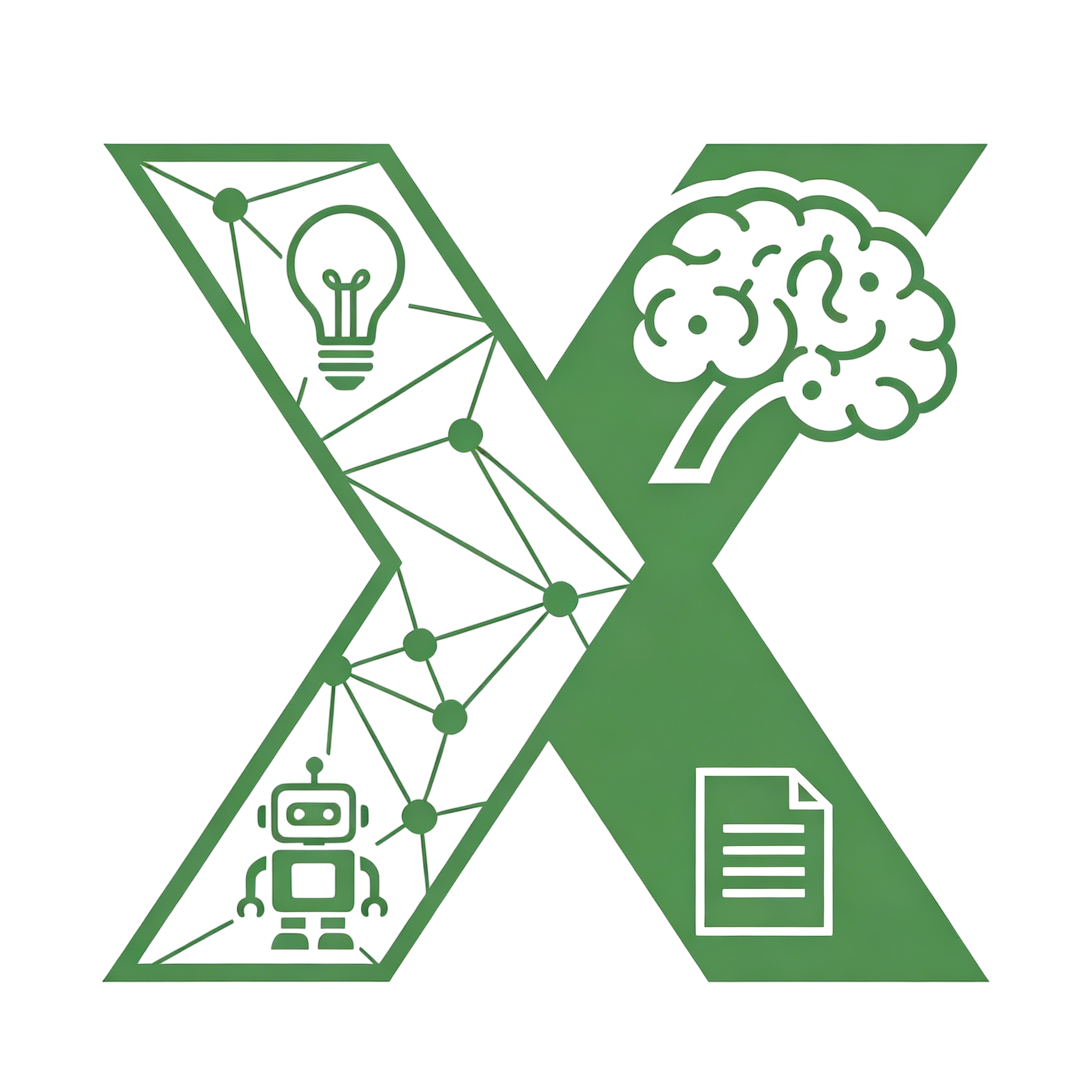}};
      \node[anchor=north east, at=(frame.north east), xshift=-0.5cm, yshift=-0.5cm]
      {\includegraphics[height=1cm]{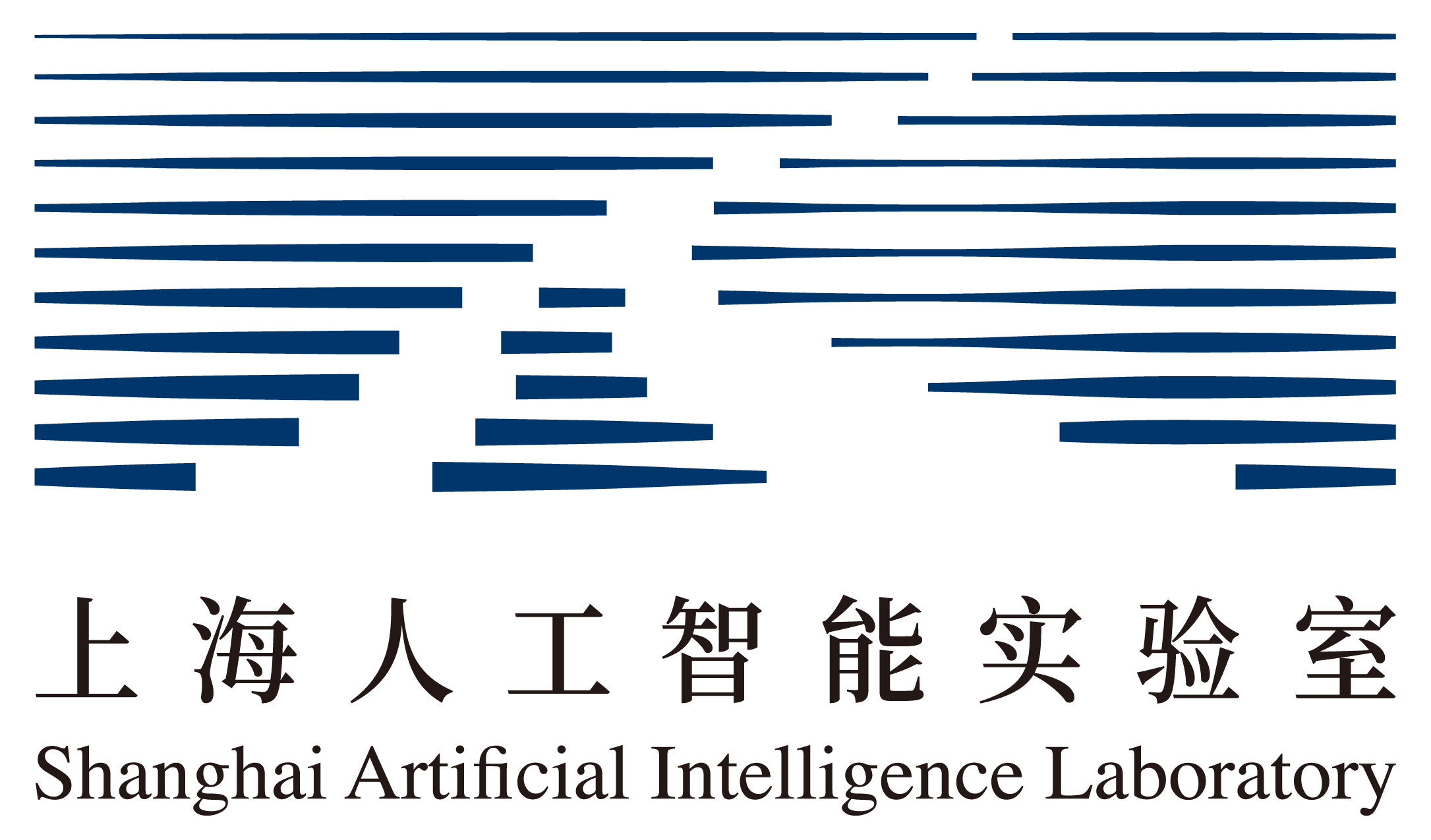}};
    }
  }%
  \begin{tcolorbox}
    \setlength{\parindent}{0cm}
    \setlength{\parskip}{0.5cm}
    {
      \setlength{\parskip}{0cm}
      \raggedright
      \nohyphens
      {
        \vskip 1.25cm 
        \setstretch{1.4}
        {\huge\sffamily\bfseries\textcolor{black}{\paperTitle}}\par
      }
      \ifshowAuthorMeta
      \vskip 0.25cm
      \paperAuthors\par
      \vskip 0.35cm
      \paperAffiliations\par
      \vskip 0.08cm
      \renderOptionalText{\paperNotes}
      \fi
    }
    \vskip 0.2cm
    {\color{textgray}%
  \begingroup
  \renewenvironment{abstract}{}{}%
  \begin{abstract}

  Existing computer-use agents remain fundamentally limited in professional software manipulation: GUI-based agents suffer from fragile visual grounding and long-horizon error accumulation, while API-based approaches struggle with heterogeneous protocols and inaccessible commercial interfaces. In this work, we identify the Component Object Model (COM) as a unified executable abstraction, proposing COM-as-Action—a new paradigm that reframes professional software interaction as deterministic program synthesis rather than sequential visual control.
  To validate this paradigm in the most demanding environments, we introduce \bench{}, the first benchmark for agents operating real industrial CAD software. Our experiments reveal a substantial paradigm gap: frontier proprietary models achieve near-zero success under GUI-based interaction, whereas COM-based execution yields substantial immediate gains. To bridge the remaining gap between syntactic correctness and geometric accuracy, we develop \actor{}, a self-correcting agent trained through a progressive three-stage framework, alongside ComForge, a scalable platform for large-scale training in Windows containers.
  Extensive experiments show that \actor{} achieves state-of-the-art performance on \bench{}, with strong resilience in long-horizon tasks where baselines collapse, and generalizes to external CAD benchmarks.

\end{abstract}
  \endgroup
\par}
    \vskip 0.4cm
    {
      \setlength{\parskip}{0cm}
      {\small {\sffamily\bfseries \raisebox{-0.2em}{\includegraphics[width=0.025\linewidth]{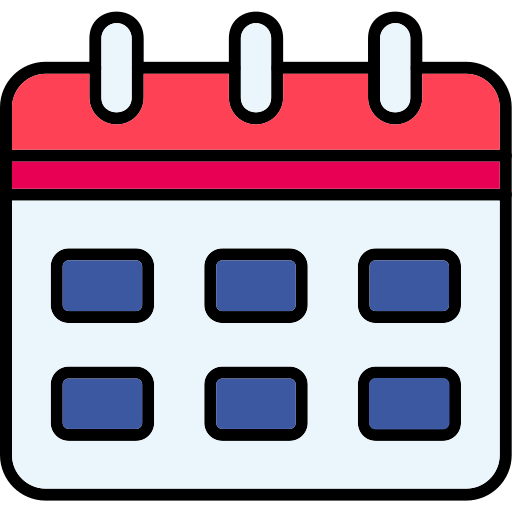}}~~Date:} \publishDate}\par%
      \vskip 0.08cm%
      {\small {\sffamily\bfseries \raisebox{-0.2em}{\includegraphics[width=0.025\linewidth]{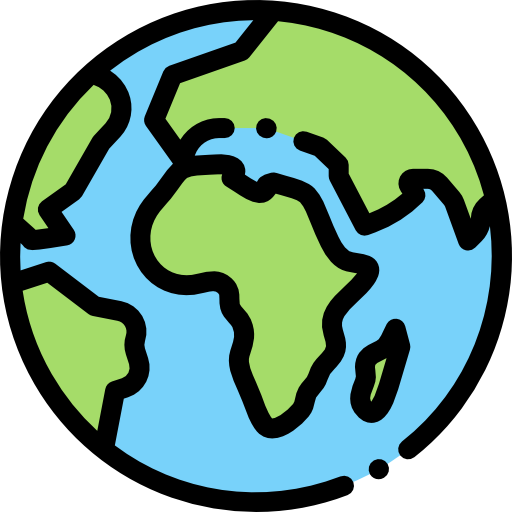}}~~Project Page:} \url{\projectURL}}\par%
      \vskip 0.08cm%
      {\small {\sffamily\bfseries \raisebox{-0.2em}{\includegraphics[width=0.025\linewidth]{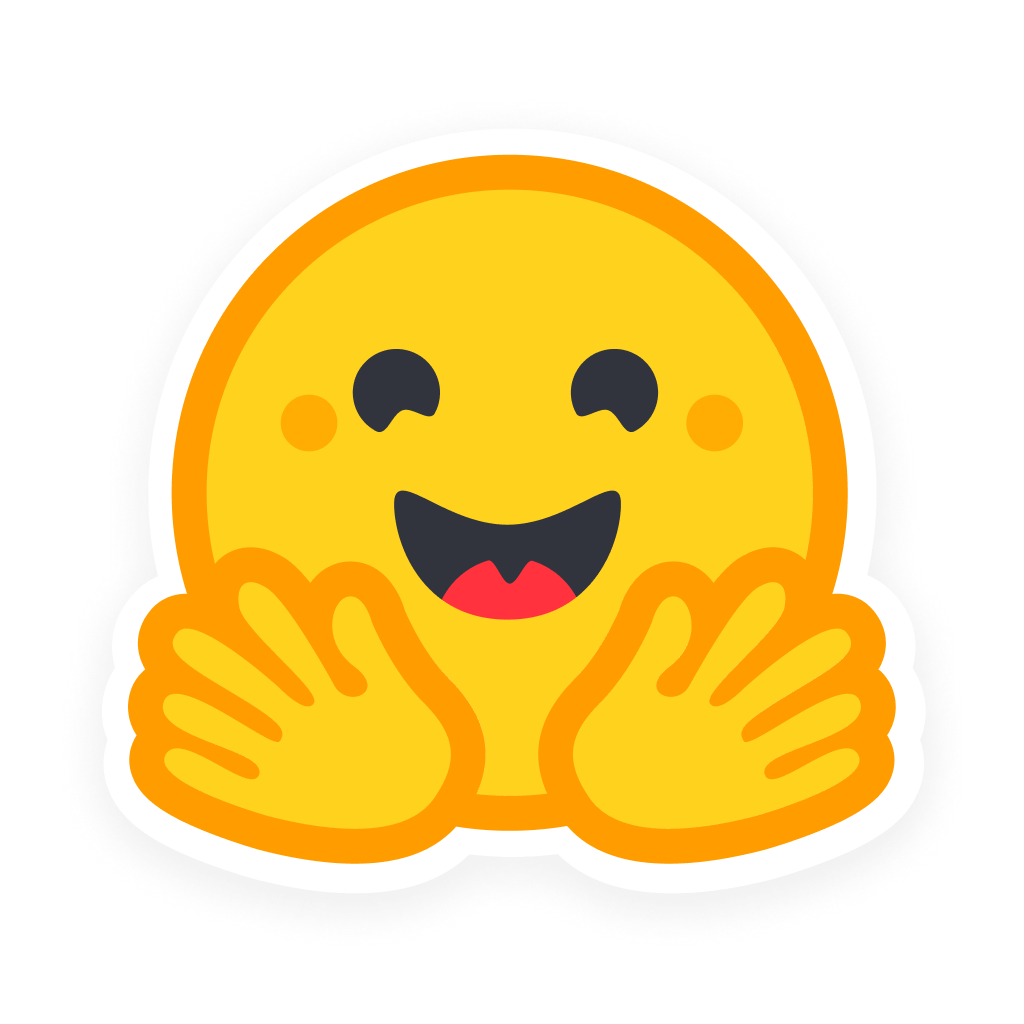}}~~Huggingface Model:} \url{\githubURL}}\par%
    }
  \end{tcolorbox}
  \tcbset{reset}
}

\begin{document}

\newgeometry{top=1in, bottom=0.75in, textwidth=6.3in, textheight=9in}
\renderFrontBox

\begin{figure}[!t]
  \centering
  \includegraphics[width=0.7\linewidth]{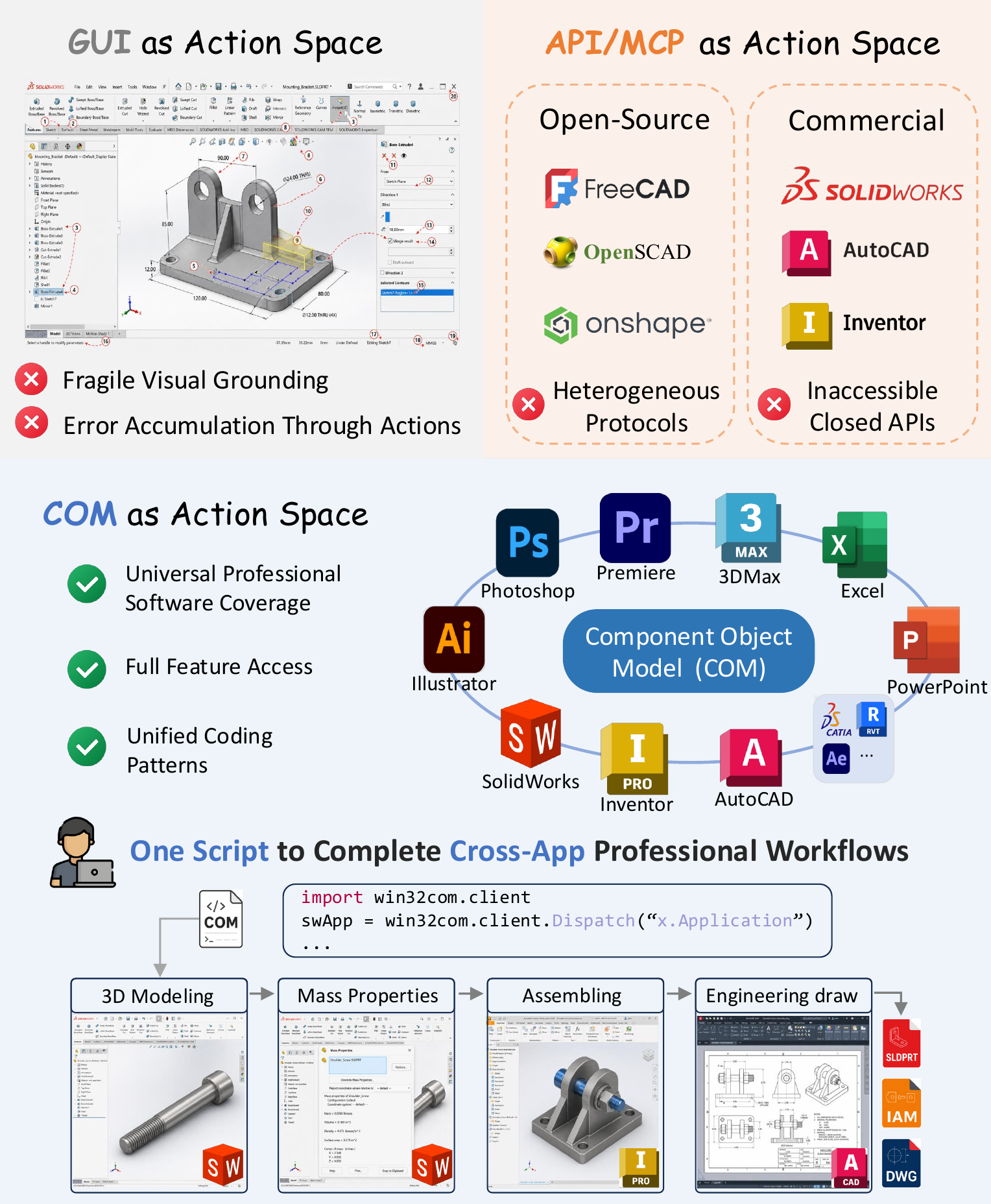}
  \caption{
    Comparison of existing computer-use paradigms and our proposed \method{} paradigm. GUI-based agents rely on fragile visual grounding and suffer from long-horizon error accumulation, while API-based agents are constrained by fragmented and limited interfaces. In contrast, we leverage the COM as a unified semantic programmatic interface, enabling executable program synthesis and cross-application workflows.
  }
  \label{fig:motivation}
  \vspace{-5pt}
\end{figure}



\section{Introduction}

Computer-use agents have recently made remarkable progress, enabling AI systems to interact with software as humans do and automate complex digital tasks across diverse applications~\cite{ref:claudecua,ref:operator}. Existing agents mainly follow two paradigms: graphical user interface (GUI)-based interaction~\cite{ref:operator} and application programming interface (API) / Model Context Protocol (MCP)-based tool invocation~\cite{ref:toolllm}. However, their application to professional software remains limited.  GUI-based agents offer broad accessibility but rely on fine-grained visual grounding, which is unreliable in visually dense professional graphic interfaces~\cite{ref:screenspot-pro}. API/MCP-based agents provide structured programmatic control, but are often constrained by fragmented, application-specific interfaces and limited public API support in commercial software. As a result, existing paradigms struggle to provide both the reliability and universality required for real-world professional workflows.

Meanwhile, MLLMs have demonstrated extraordinary proficiency in code generation~\cite{ref:gpt-5}, achieving state-of-the-art performance on benchmarks such as SWE-bench~\cite{ref:swebench} and Codeforces~\cite{ref:livecodebench}. This raises a compelling question: can agents leverage this code synthesis capacity to manipulate professional software programmatically, bypassing both the visual fragility of GUI interactions and the tool scarcity of closed-source applications? We identify the Component Object Model (COM)~\cite{ref:com}—the native interoperability framework widely supported across the Windows ecosystem—as a natural answer. Unlike accessibility-oriented frameworks such as UI Automation (UIA)~\cite{ref:osworld},
COM operates at the semantic object level, exposing application internals through structured programmatic interfaces across professional software including Microsoft Office and SolidWorks.

Building on this insight, we propose COM-as-Action (\method{}), a new paradigm that reframes professional software manipulation as executable program synthesis: agents directly manipulate semantic software objects through generated COM programs, while reserving visual perception for coarse-grained observation and verification.

\method{} resolves both core bottlenecks of professional software automation. First, it addresses the tool deficit in closed-source software: while industry leaders such as Microsoft, Autodesk, and Adobe conceal their source code, they natively expose comprehensive COM interfaces, giving agents deep programmatic access as a high-fidelity local API. Second, code-driven execution eliminates cascading errors in long-horizon tasks—rather than performing hundreds of sequential GUI interactions, the agent generates a single deterministic script, shifting from a vulnerable executor to a high-level programmer. Furthermore, unlike existing code-based agents that rely on fragmented, application-specific APIs, \method{} consolidates a unified action space spanning Microsoft Office, Adobe suites, and industrial CAD applications, naturally supporting cross-application workflows.

While COM interfaces can be applied to a wide range of professional fields, we focus on industrial Computer-Aided Design (CAD) as the primary validation domain. CAD software embodies the core bottlenecks of digital automation: its cluttered interfaces demand high-precision 3D spatial reasoning, while completing a model requires hundreds of highly correlated geometric operations that readily expose the fragility of traditional GUI agents. Furthermore, authentic CAD workflows naturally entail complex, cross-platform task sequences, providing an ideal testbed for evaluation. Overcoming the stringent accuracy and long-horizon operational requirements of CAD engineering will strongly validate the broader applicability of our framework across other professional fields.

To operationalize and evaluate this paradigm, we introduce a complete ecosystem for code-driven CAD software automation. First, we establish \bench{}, the first benchmark designed to evaluate agents operating via real CAD COM interfaces (e.g., AutoCAD and SolidWorks) based on final software artifacts. Preliminary experiments show that traditional GUI agents achieve near-zero success rates, whereas replacing GUI interactions with COM-based actions leads to substantial immediate improvements. Nevertheless, few-shot prompting alone remains far from reliable. To bridge this gap, we develop \actor{}, an autonomous agent that utilizes COM as its action space to translate instructions into Python scripts, iteratively refining them through real-time screenshot and terminal feedback. To scale this paradigm, we build ComForge, a highly parallelized infrastructure hosting thousands of concurrent, Dockerized Windows environments. Leveraging this platform, we propose a progressive three-stage training framework—incorporating single-turn SFT, multi-turn SFT, and Reinforcement Learning (RL) guided by a continuous geometric reward—to evolve \actor{} from a static code generator into a self-correcting agent. Driven by this framework, \actor{} achieves state-of-the-art performance on \bench{} and demonstrates exceptional generalization across external CAD benchmarks.

Our contributions can be summarized as follows:
\begin{itemize}[leftmargin=*, itemindent=0pt, itemsep=0pt, topsep=2pt]
  \item We propose \method{}, a new computer-use paradigm, replacing fragile GUI interaction with executable COM program synthesis, establish a unified action space for professional software.
  \item We build ComForge, a scalable platform for large-scale training in real computer environments, and develop \actor{}, a professional CAD agent, through a progressive three-stage framework.
  \item We introduce \bench{}, the first benchmark for evaluating agents in real CAD design environments via real CAD COM interfaces, and demonstrate through extensive experiments that \actor{} achieves SOTA performance on \bench{} while generalizing to external CAD benchmarks including Text2CAD and CADPrompt.

\end{itemize}


\section{Related Works}
\subsection{Computer-use Agents}
Existing computer-use agents mainly follow two paradigms: GUI-based and API-based interaction.
GUI-based agents~\cite{ref:operator,ref:seeclick,ref:agents,ref:uitars,ref:ufo3,ref:cogagent,ref:appagent,ref:aria,ref:gpt4viswebagent}, powered with MLLMs, rely on screenshot-based visual grounding and low-level mouse-and-keyboard control, achieving strong performance on general desktop tasks but remaining fragile in visually dense professional software environments. Programmatic agents~\cite{ref:toolllm,ref:claudecua,ref:llmbrainedguiagents,ref:osagents,ref:beyondbrowsing,ref:autowebglm,ref:oscopilot,ref:executablecodeactions,ref:coact} instead execute commands or APIs directly, providing more reliable execution but relying on fragmented, application-specific interfaces with limited interoperability across software ecosystems. In contrast, our work introduces COM as a unified, code-centric action space that combines the reliability of programmatic execution with the broad functional coverage of GUI interaction, while reserving visual perception for coarse-grained observation and verification.

\subsection{CAD Generation}

Existing CAD generation methods mainly fall into two categories: sequence generation and code generation. Sequence-based approaches~\cite{ref:sketchgraphs,ref:fusion360gallery,ref:transcad,ref:cadllama,ref:flexcad} such as DeepCAD~\cite{ref:deepcad} and Text2CAD~\cite{ref:text2cad} model CAD construction as parametric command sequences, while code-based approaches including CADCoder~\cite{ref:cadcoder} and CAD-Recode~\cite{ref:cadrecode} generate executable CAD scripts in libraries such as CadQuery and build123d. Although these methods demonstrate promising capabilities for geometric modeling, they remain limited to constrained operation vocabularies or lightweight geometric libraries, lacking support for complex industrial workflows and advanced functionalities in professional CAD software. In contrast, our work directly treats industrial CAD software as the execution environment through COM-based interaction, enabling agents to manipulate commercial CAD applications and operate across diverse engineering workflows.

\section{Preliminary}

\begin{figure*}[!t]
  \centering
  \includegraphics[width=\linewidth]{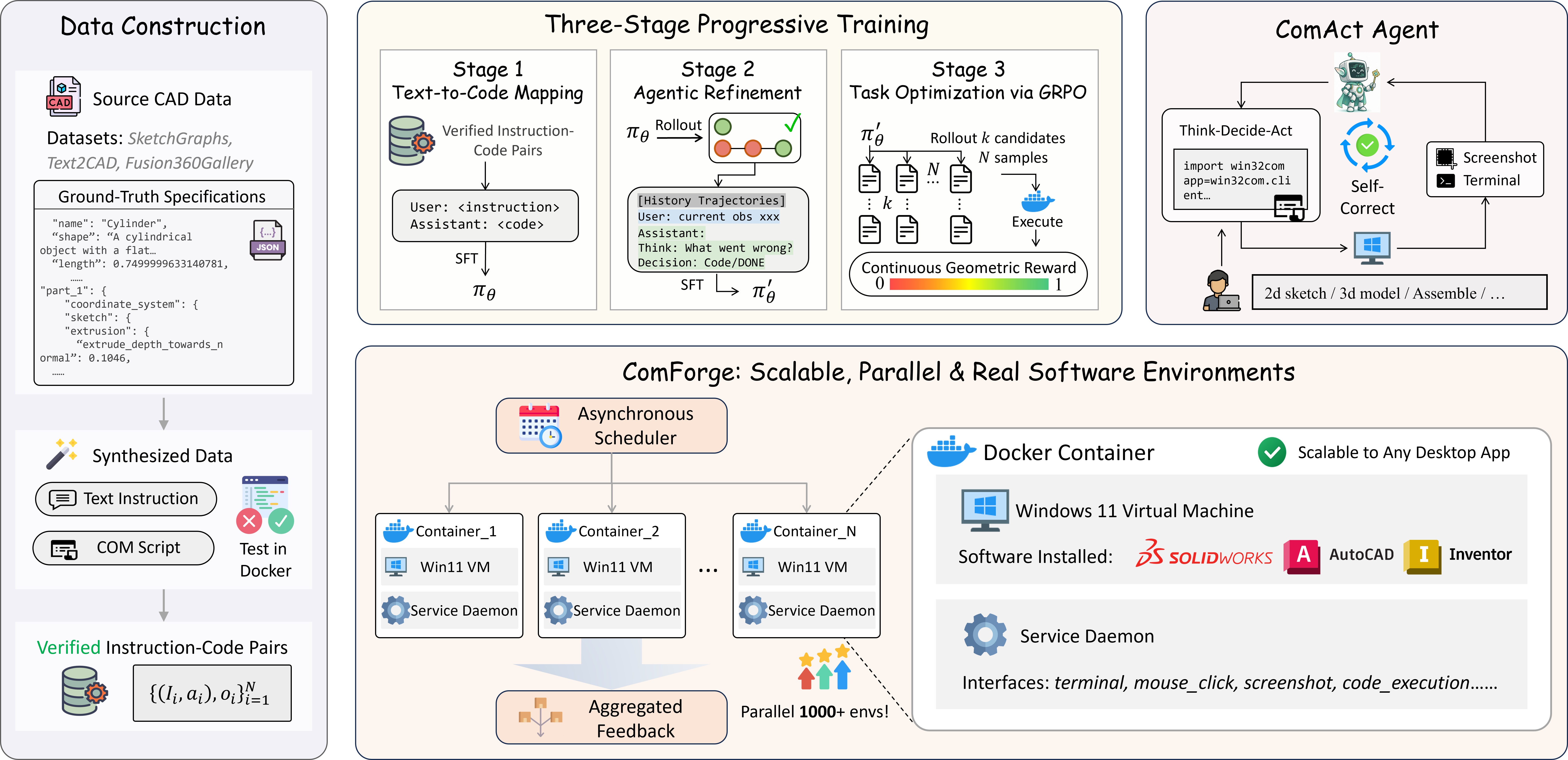}
  \vspace{-15pt}
  \caption{
    Overview of our ComAct framework, consisting of three components: a data construction pipeline that synthesizes verified instruction–code pairs; a three-stage progressive training framework (text-to-code SFT, agentic SFT, and GRPO with continuous geometric reward); and a scalable infrastructure supporting 1000+ parallel real Windows environments for training and evaluation.
  }
  \label{fig:pipeline}
  \vspace{-10pt}
\end{figure*}

\subsection{Component Object Model}



The Component Object Model (COM) is a binary interface standard introduced by Microsoft that enables programmatic communication between desktop applications. A COM-enabled application exposes its functionality as a hierarchy of objects with callable properties and methods, which can be accessed from languages with COM bindings such as Python via \texttt{win32com.client}.
Unlike application-specific scripting APIs, COM provides a unified system-level interface broadly supported across professional software ecosystems including AutoCAD, SolidWorks, Office, and Adobe applications. Operations accessible through the GUI can be executed directly through COM calls, while native cross-application communication enables workflows spanning multiple software tools.
This combination of completeness, uniformity, and interoperability allows COM to serve as a unified action space for professional software manipulation. A short COM script can replace dozens of sequential GUI actions, transforming software interaction from low-level visual control into executable program synthesis.


\subsection{Formulation of COM as Action}

We formalize professional software manipulation as a partially observable Markov decision process $\langle \mathcal{S}, \mathcal{O}, \mathcal{A}, \mathcal{T}, \Omega \rangle$, where $\mathcal{S}, \mathcal{O}, \mathcal{A}$ denote the state, observation, and action spaces, $\mathcal{T}: \mathcal{S} \times \mathcal{A} \to \mathcal{S}$ is the transition function, and $\Omega: \mathcal{S} \to \mathcal{O}$ is the observation function. Crucially, each action $a \in \mathcal{A}$ is an executable Python script invoking COM interfaces rather than atomic GUI operations. Given a natural language instruction $\mathcal{I}$ and initial state $s_0$, the agent iteratively receives an observation $o_t = \Omega(s_t) \in \mathcal{O}$ (screenshot and execution feedback) and predicts a decision token $\delta_t \in \{\texttt{CODE}, \texttt{DONE}, \texttt{FAIL}\}$. If $\delta_t = \texttt{CODE}$, the agent generates and executes a script $a_t \in \mathcal{A}$ to transition the state; otherwise, the loop terminates. The resulting trajectory is evaluated by whether the final software artifact satisfies $\mathcal{I}$.

\section{Methodology}


Figure~\ref{fig:pipeline} overviews our framework. We first construct a verified instruction–code corpus via automated synthesis (Sec.~\ref{sec:data}). Using this data, we train \actor{} through a progressive three-stage pipeline spanning SFT and GRPO optimization (Sec.~\ref{sec:agent}). This process is anchored in ComForge, a scalable infrastructure hosting parallelized, Dockerized Windows environments with real CAD software (Sec.~\ref{sec:forge}). At inference, the agent executes a closed-loop ``Think–Decide–Act'' cycle, generating and refining COM-based Python scripts via real-time environment feedback.

\subsection{Training Data Construction}
\label{sec:data}

We construct a corpus of verified instruction–code pairs from three public CAD datasets: SketchGraphs~\cite{ref:sketchgraphs}, Text2CAD~\cite{ref:text2cad}, and Fusion360Gallery~\cite{ref:fusion360gallery}. Each sample provides a ground-truth geometry specification in structured JSON format, encoding object geometry, sketch constraints, and feature operations.
From each JSON file, we synthesize a training sample in two steps. First, we prompt an MLLM to generate a natural language instruction describing the target geometry, and a corresponding COM script that reproduces the model. Second, each generated script is executed inside an isolated Docker container running a real Windows environment with CAD software installed.  Only samples that pass this validation are retained.  We ensure that the SFT corpus contains only syntactically and semantically correct programs, providing a reliable foundation for subsequent training stages.

\vspace{-10pt}
\subsection{ComAct Agent: A Self-Correcting Agent for Professional Software}
\vspace{-10pt}
\label{sec:agent}
\noindent \textbf{Overview.}
Building upon the ComAct paradigm and the ComForge platform, we train ComActor, a multi-modal agent that translates natural language instructions into executable COM scripts and iteratively self-corrects through continuous environment interaction until task completion.
ComActor operates as a closed-loop policy $\pi_\theta$ within ComForge, following the multi-turn interaction protocol defined in ComAct.
We train $\pi_\theta$ in a progressive three-stage framework, with each stage targeting a specific capability bottleneck exposed by the previous one, collectively evolving the model from a static COM code generator into a self-correcting agent that reliably executes long-horizon professional workflows.

\noindent \textbf{Stage 1: Instruction-to-Code Mapping.}
Off-the-shelf MLLMs lack inherent familiarity with COM APIs, frequently producing syntax errors or hallucinating non-existent API calls. To bootstrap a base policy, we perform supervised fine-tuning on the verified instruction–code corpus $\mathcal{D}_{\text{S1}} = \{(\mathcal{I}_i, a_i)\}_{i=1}^N$ constructed in Sec.\ref{sec:data}, with the standard autoregressive objective $\mathcal{L}_{\text{S1}}(\theta) $:
\begin{equation}
  \small
  - \mathbb{E}_{(\mathcal{I}, a_{\text{gt}}) \sim \mathcal{D}_{\text{S1}}} \left[ \sum_{k=1}^{|a_{\text{gt}}|} \log \pi_\theta \left(a^{(k)} \mid a^{(<k)}, \mathcal{I}\right) \right]
\end{equation}
Stage 1 equips $\pi_\theta$ with the basic capacity to generate syntactically valid COM scripts, but the resulting static policy has no mechanism for recovering from execution failures.

\noindent \textbf{Stage 2: Agentic Refinement with Multimodal Feedback.}
Stage 1 yields a static policy with no mechanism for recovering from execution failures. To address this, we fine-tune $\pi_\theta$ on multi-turn interaction trajectories collected by rolling out the Stage-1 policy in ComForge on a held-out instruction set. Successful trajectories are retained directly as positive supervision; for failed ones, a strong teacher model diagnoses the errors and rewrites the reasoning and code into corrected trajectories. The resulting dataset $\mathcal{D}_{\text{S2}}$ covers diverse interaction patterns including error diagnosis, iterative code revision, and task termination. We optimize $\pi_\theta$ with the following objective $\mathcal{L}_{\text{S2}}(\theta)$:
\begin{equation}
  \small
  \begin{split}
    - \mathbb{E}_{\tau \sim \mathcal{D}_{\text{S2}}} \Bigg[ \sum_{t=0}^{|\tau|} \sum_{k=1}^{|\mathcal{Z}_t|}  \log \pi_\theta \Big( \mathcal{Z}_t^{(k)} \mid \mathcal{Z}_t^{(<k)}, o_{\leq t}, \mathcal{Z}_{<t}, \mathcal{I} \Big) \Bigg]
  \end{split}
\end{equation}
where  $o_t$ represents the environment feedback, $r_t$ denotes the reasoning trace, $\delta_t$ denotes the decision token and $a_t$ is the predicted COM code, $\mathcal{Z}_t^{(k)}=(r_t, \delta_t, a_t)^{(k)}$ denotes the $k$-th token of the target sequence at step $t$; $o_{\leq t}$ and $\mathcal{Z}_{<t}=(r, \delta, a)_{<t}$ represent the accumulated interaction history.

After Stage 2, the agent can interpret feedback and iteratively debug its code. Yet a subtler failure mode remains: error-free code may still produce artifacts that diverge from the task specification, motivating task-level optimization in Stage 3.

\noindent \textbf{Stage 3: Task-Level Optimization via GRPO.}
To close the gap between code-level correctness and task-level fidelity, we optimize the agent directly with on-policy RL using GRPO~\cite{shao2024deepseekmath}.

\noindent\textbf{Continuous Geometric Reward.}
We measure geometric fidelity via Chamfer Distance (CD) between normalized point clouds of the generated artifact and the ground truth, sampled from mesh surfaces (3D STL) or parametric curves (2D DXF). To avoid reward sparsity, we convert CD into a continuous reward via log-linear decay:
\begin{equation}
  \small
  \begin{split}
    R_{\text{geo}}(\mathrm{CD}) &= \mathbb{I}(\mathrm{CD} \leq c_{\text{low}}) \quad + \mathbb{I}(c_{\text{low}} < \mathrm{CD} \leq c_{\text{high}}) \\ & \cdot \left(1 + \frac{\log\mathrm{CD} - \log c_{\text{low}}}{\log c_{\text{high}} - \log c_{\text{low}}} \cdot (r_{\min} - 1)\right)
  \end{split}
\end{equation}
where $c_{\text{low}}, c_{\text{high}}$ are the near-perfect and failure thresholds, and $r_{\min}=0.01$. Execution failures receive $R_{\text{geo}}=0$.

\noindent\textbf{GRPO Optimization.}
We curate $\mathcal{D}_{\text{S3}}$ from ComForge rollouts of the Stage-2 policy, retaining instructions where execution succeeds yet task verification fails. For each $\mathcal{I} \in \mathcal{D}_{\text{S3}}$, we sample $K$ candidate scripts, execute them to obtain rewards $\{R_i\}$, and compute group-normalized advantages $\hat{A}_i = (R_i - \mu_R)/\sigma_R$. The GRPO objective is:
\begin{equation}
  \small
  \begin{split}
    \mathcal{L}_{\text{GRPO}}(\theta) &= - \frac{1}{K} \sum_{i=1}^K \frac{1}{|a_i|} \sum_{t=1}^{|a_i|} \min\!\Big( \rho_{i,t} \hat{A}_i,\; \\
    &\mathrm{clip}(\rho_{i,t}, 1{\pm}\epsilon)\, \hat{A}_i \Big) \quad + \beta\, D_{\text{KL}}(\pi_\theta \| \pi_{\text{ref}})
  \end{split}
\end{equation}
where $\rho_{i,t} = \pi_\theta / \pi_{\theta_{\text{old}}}$ is the token-level probability ratio, $\epsilon$ is the clipping hyperparameter, and $\beta$ controls the KL penalty.

\subsection{ComForge: Scalable Parallel Environments for CAD Software}
\label{sec:forge}

Training agents for professional software requires execution within authentic environments at massive scale—a need that existing benchmarks, which rely on static datasets or lightweight sandboxes, cannot meet. We build ComForge, a scalable platform of real computer environments that is architecturally general and extensible to any desktop software.

\noindent \textbf{Containerized Environments and Unified Interface.}
Each ComForge environment is a Docker container hosting a full Windows virtual machine with target applications pre-installed. A lightweight service daemon within each container exposes a unified interface of three action types: \texttt{code\_execute} (running COM-based Python scripts), \texttt{gui\_action} (mouse and keyboard events), and \texttt{reset} (restoring initial state). After each action, the daemon returns a screenshot and terminal output as the observation, abstracting disparate software instances into a homogeneous request-response interface.

\noindent \textbf{Massive Parallelization and Training Integration.}
Since professional software operations incur non-trivial execution times, serial collection would bottleneck RL training. ComForge addresses this with an asynchronous scheduler that dynamically routes batched requests to idle containers and aggregates observations in parallel, scaling linearly to thousands of concurrent environments. We integrate ComForge into the ms-swift GRPO pipeline, where it handles action execution and reward computation across all three training stages.

\begin{figure}[!t]
  \centering
  \includegraphics[width=0.7\linewidth]{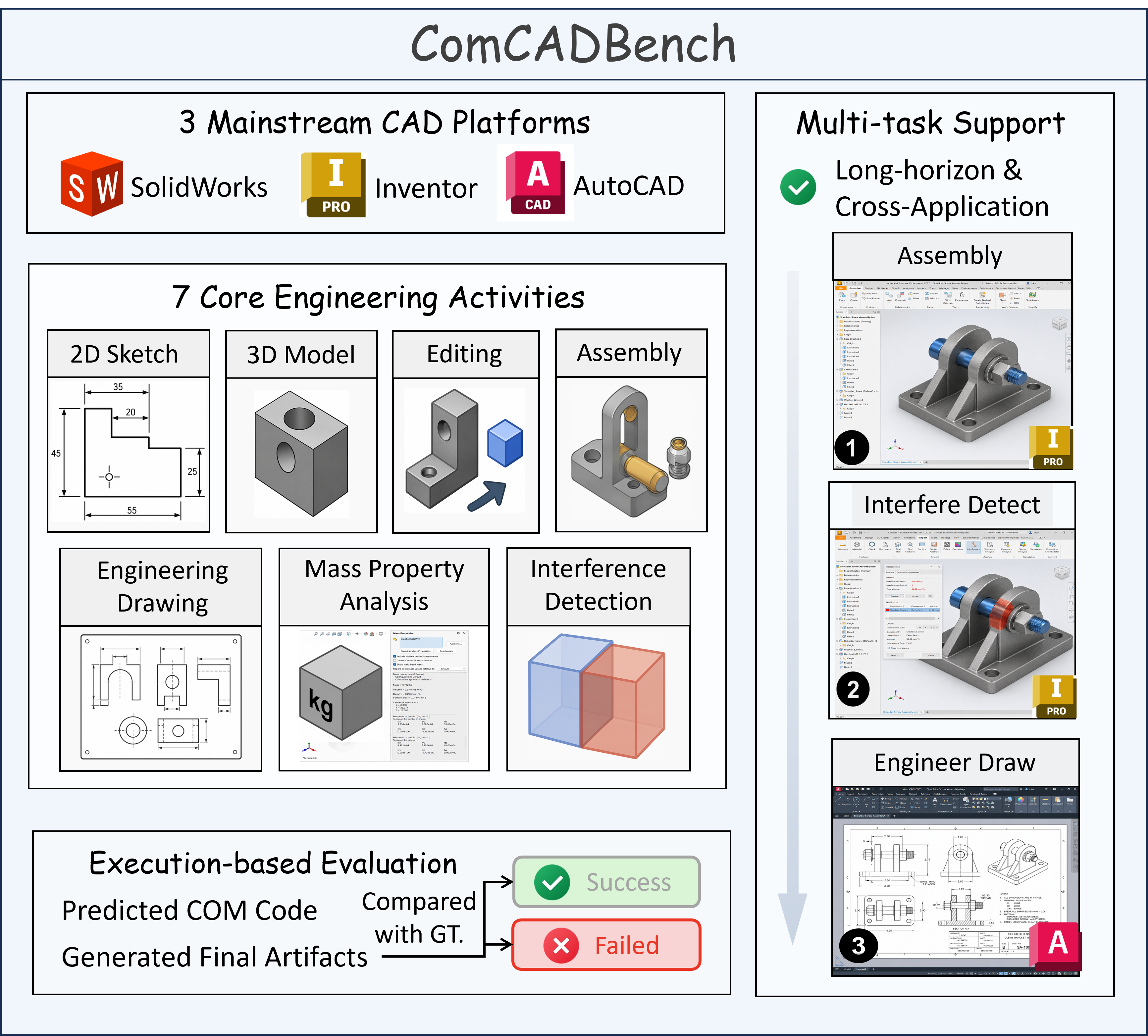}
  \caption{
    ComCADBench covers 3 CAD platforms, 7 engineering activities, and supports long-horizon cross-application tasks, evaluated by comparing generated artifacts against ground truth.
  }
  \label{fig:bench}
  \vspace{-10pt}
\end{figure}




\begin{table*}[t]
  \centering
  \scriptsize
  \setlength{\tabcolsep}{3.5pt}
  \renewcommand{\arraystretch}{1.1}

  \resizebox{\textwidth}{!}{%
    \begin{tabular}{llcccccccccc}
      \toprule

      \multicolumn{2}{c}{\multirow{4.5}{*}{\shortstack{\textbf{Tasks} \\ \textbf{VR / SR}\,$\uparrow$}}}
      & \multicolumn{4}{c}{\textbf{Single-Task (400)}}
      & \multicolumn{6}{c}{\textbf{Multi-Task (600)}} \\

      \cmidrule(lr){3-6}
      \cmidrule(lr){7-12}

      \multicolumn{2}{c}{}
      & \multicolumn{2}{c}{\textbf{3D-M (200)}}
      & \textbf{Asm. (100)}
      & \textbf{2D-S (100)}
      & \multicolumn{2}{c}{\textbf{3D-M+E (200)}}
      & \textbf{3D-M+D (50)}
      & \multicolumn{2}{c}{\textbf{3D-M+MP (200)}}
      & \textbf{Asm.+ID} \\

      \cmidrule(lr){3-4}
      \cmidrule(lr){7-8}
      \cmidrule(lr){10-11}

      \multicolumn{2}{c}{}
      & \textbf{SW (100)}
      & \textbf{Inv. (100)}
      & \textbf{Inv. (100)}
      & \textbf{ACAD (100)}
      & \textbf{SW (100)}
      & \textbf{Inv. (100)}
      & \textbf{SW (100)}
      & \textbf{SW (100)}
      & \textbf{Inv. (100)}
      & \textbf{Inv. (100)} \\

      \midrule
      \rowcolor{gray!8}
      \multicolumn{12}{c}{\textbf{\textit{GUI as Action}}} \\
      \midrule
      \modelicon{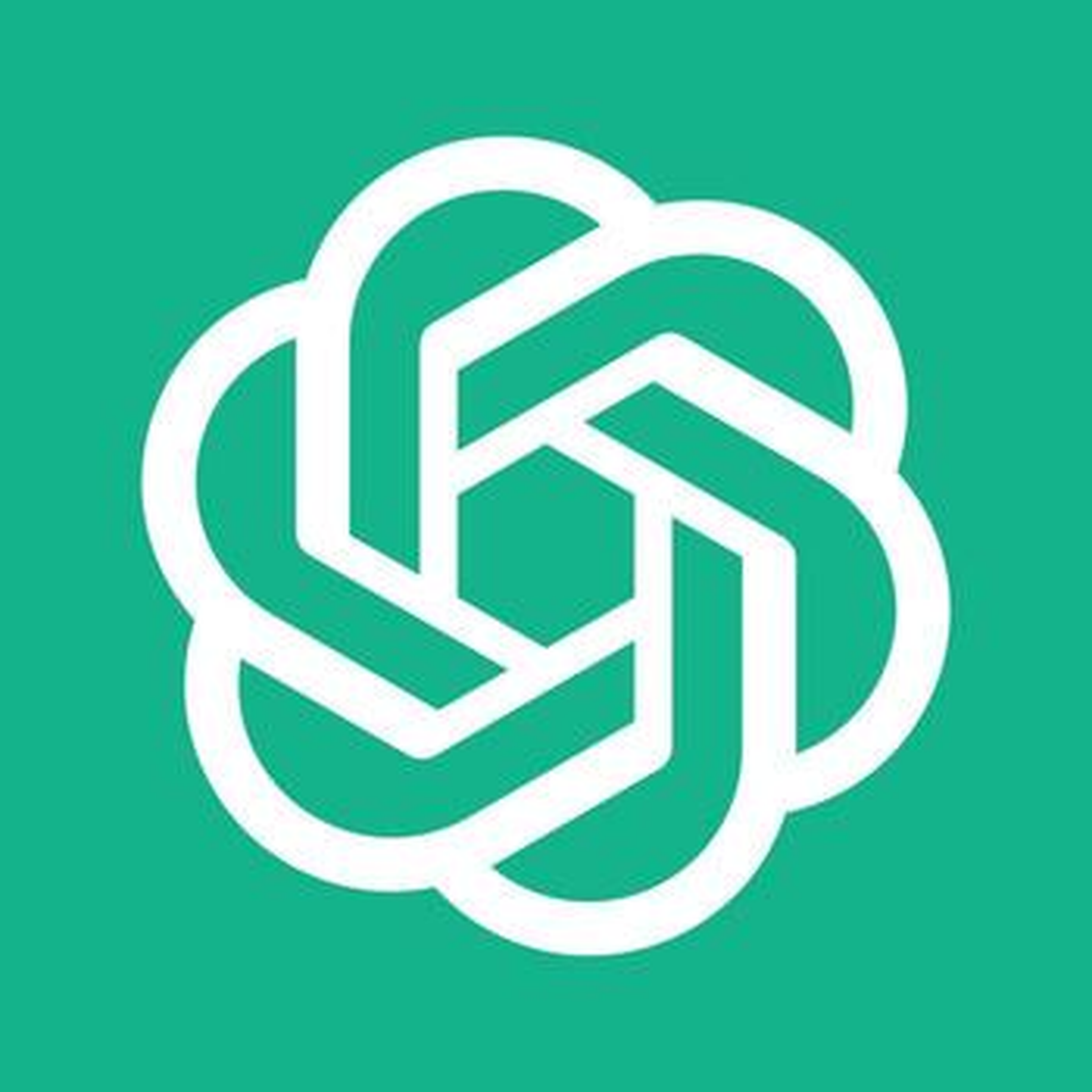}\hspace{2pt}GPT-5
      & & 0.0 / 0.0 & 0.0 / 0.0 & 0.0 / 0.0 & 0.0 / 0.0 & 0.0 / 0.0 & 0.0 / 0.0 & 0.0 / 0.0 & 0.0 / 0.0 & 0.0 / 0.0 & 0.0 / 0.0 \\
      \modelicon{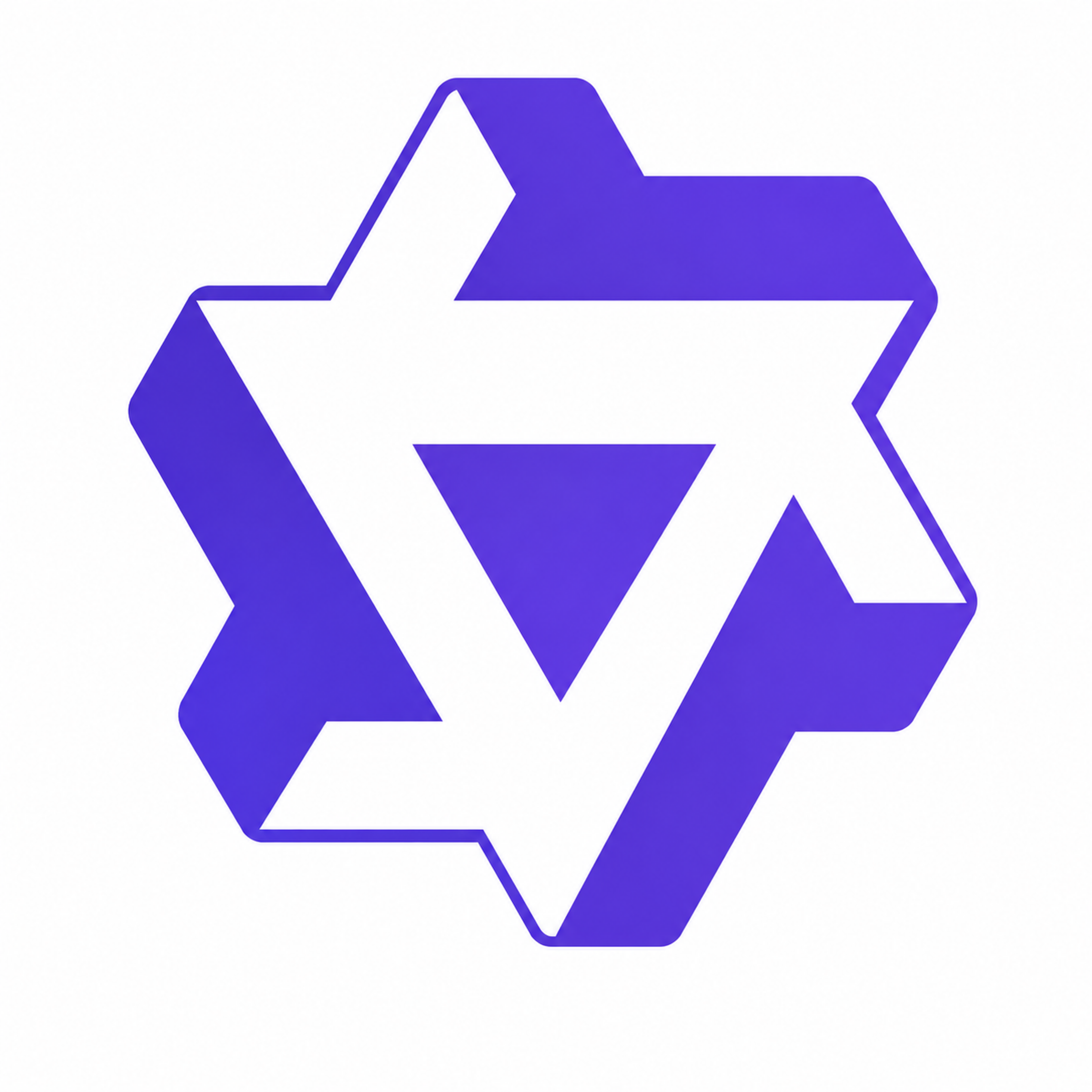}\hspace{2pt}Qwen3.5-9B
      & & 0.0 / 0.0 & 0.0 / 0.0 & 0.0 / 0.0 & 0.0 / 0.0 & 0.0 / 0.0 & 0.0 / 0.0 & 0.0 / 0.0 & 0.0 / 0.0 & 0.0 / 0.0 & 0.0 / 0.0 \\

      \midrule
      \rowcolor{gray!8}
      \multicolumn{12}{c}{\textbf{\textit{COM as Action}}} \\
      \midrule
      \modelicon{openai_logo.pdf}\hspace{2pt}GPT-5
      & & 24.0 / 7.0 & 28.0 / 21.0 & 5.0 / 2.0 & 71.0 / 33.0 & 24.0 / 7.0 & 19.0 / 15.0 & 0.0 / 0.0 & 13.0 / 6.0 & 25.0 / 10.0 & 0.0 / 0.0 \\
      \modelicon{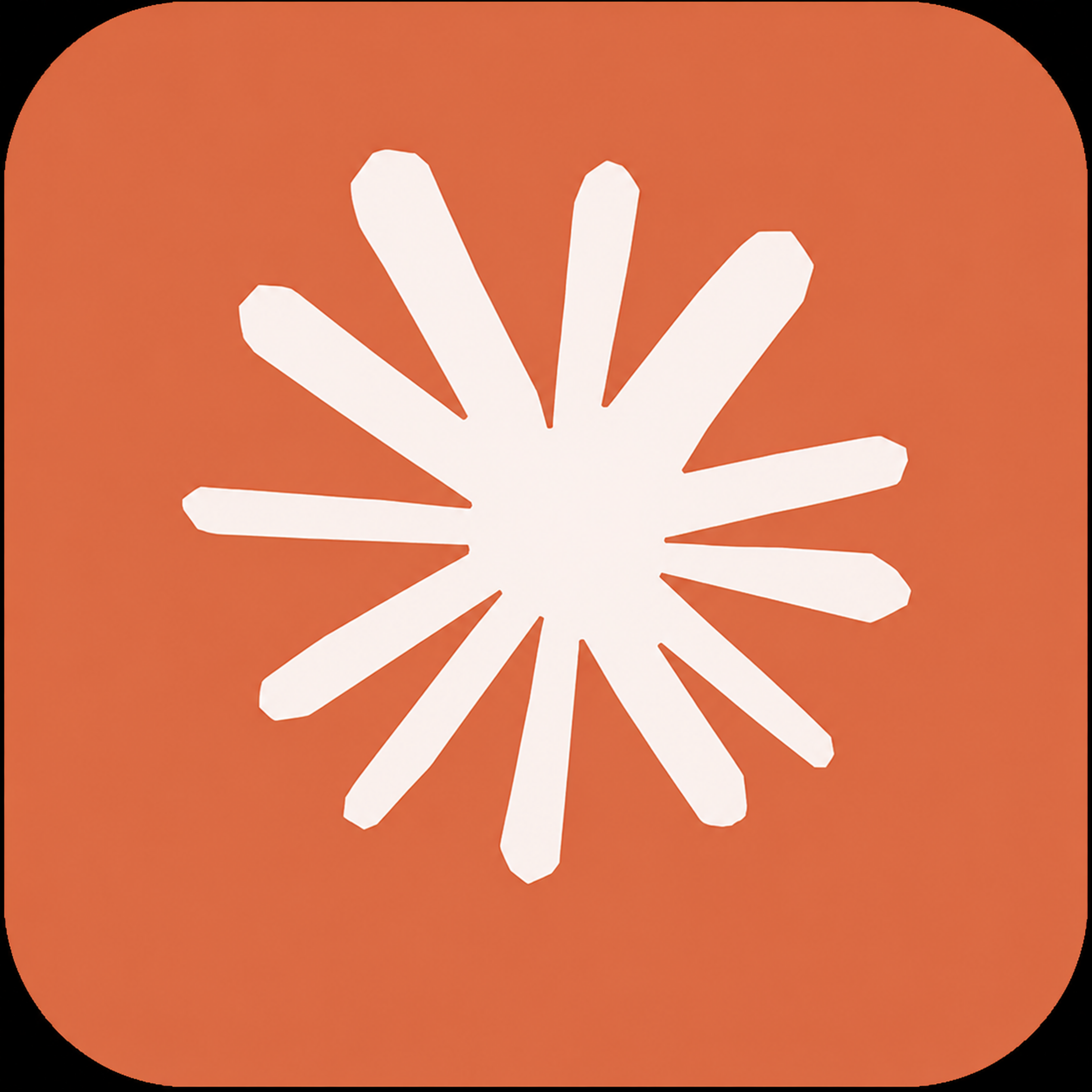}\hspace{2pt}Claude-Sonnet-4.6
      & & 7.0 / 5.0 & 8.0 / 8.0 & 3.0 / 1.0 & 63.0 / 52.0 & 4.0 / 2.0 & 4.0 / 4.0 & 0.0 / 0.0 & 0.0 / 0.0 & 0.0 / 0.0 & 0.0 / 0.0 \\
      \modelicon{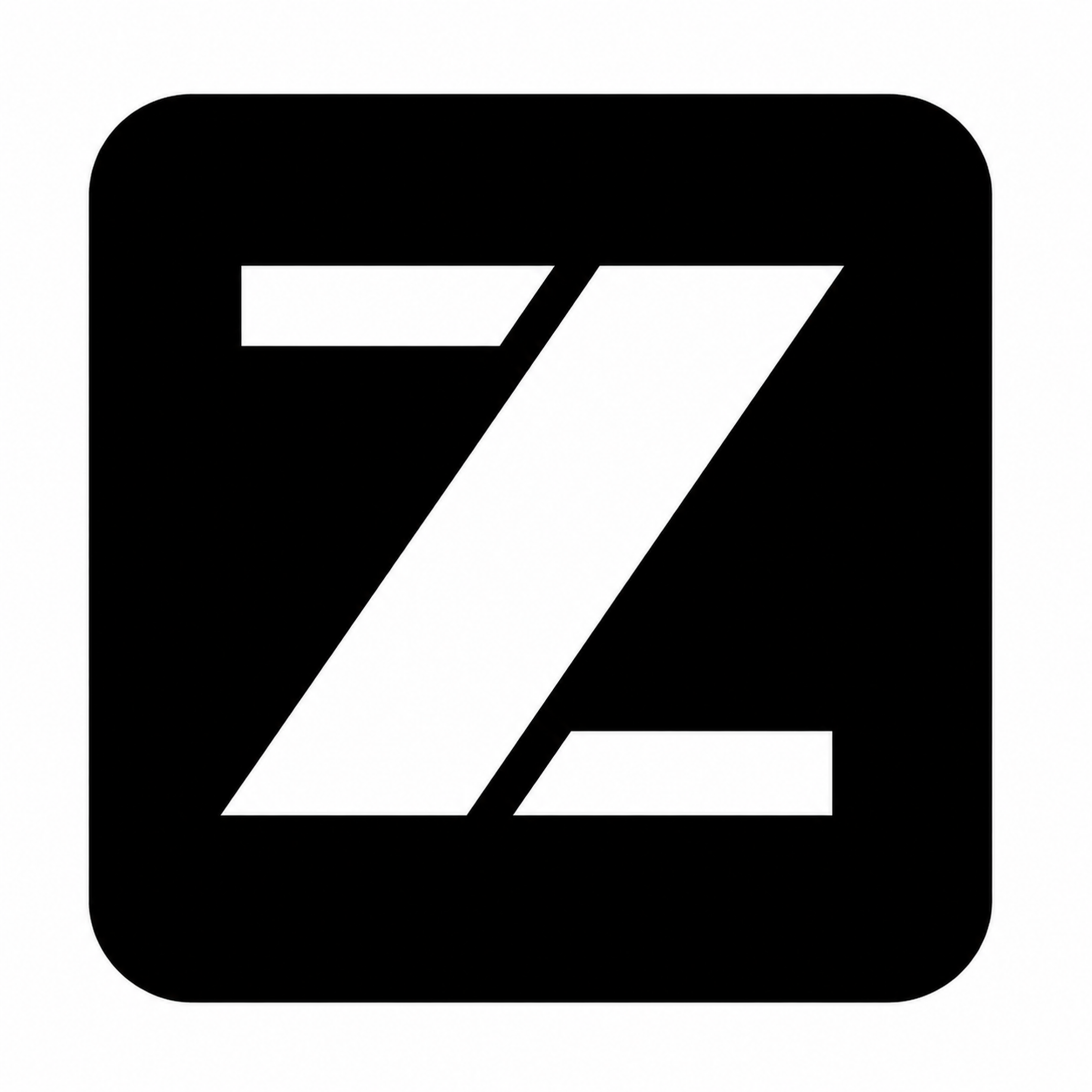}\hspace{2pt}GLM-4.6v
      & & 0.0 / 0.0 & 0.0 / 0.0 & 0.0 / 0.0 & 0.0 / 0.0 & 0.0 / 0.0 & 0.0 / 0.0 & 0.0 / 0.0 & 0.0 / 0.0 & 0.0 / 0.0 & 0.0 / 0.0 \\
      \modelicon{qwen_logo.pdf}\hspace{2pt}Qwen3.5-35B
      & & 0.0 / 0.0 & 0.0 / 0.0 & 0.0 / 0.0 & 0.0 / 0.0 & 0.0 / 0.0 & 0.0 / 0.0 & 0.0 / 0.0 & 0.0 / 0.0 & 0.0 / 0.0 & 0.0 / 0.0 \\
      \modelicon{qwen_logo.pdf}\hspace{2pt}Qwen3.5-9B
      & & 0.0 / 0.0 & 0.0 / 0.0 & 0.0 / 0.0 & 0.0 / 0.0 & 0.0 / 0.0 & 0.0 / 0.0 & 0.0 / 0.0 & 0.0 / 0.0 & 0.0 / 0.0 & 0.0 / 0.0 \\

      \midrule
      \rowcolor{gray!8}
      \multicolumn{12}{c}{\textbf{\textit{COM as Action with Few-Shot Prompting}}} \\
      \midrule
      \modelicon{openai_logo.pdf}\hspace{2pt}GPT-5
      & & 64.0 / 36.0 & 69.0 / 60.0 & 67.0 / \underline{56.0} & \underline{98.0} / 58.0 & 66.0 /40.0 & 54.0 /34.0 & 50.0 / \underline{43.0} & 58.0 / 40.0 & 62.0 / 26.0 & 62.0 / 42.0 \\
      \modelicon{claude_logo.pdf}\hspace{2pt}Claude-Sonnet-4.6
      & & \underline{76.0} / \underline{74.0} & \underline{82.0} / \underline{60.0} & 57.0 / 55.0 & 96.0 / \textbf{88.0} & \underline{81.0} / \textbf{77.0} & \underline{56.0} / \underline{40.0} & \underline{69.0} / 22.0 & \underline{73.0} / \underline{51.0} & \underline{72.0} / \underline{34.0} & 52.0 / 34.0 \\
      \modelicon{glm_logo.pdf}\hspace{2pt}GLM-4.6v
      & & 40.0 / 8.0 & 21.0 / 12.0 & 11.0 / 11.0 & \textbf{100.0} / 82.0 & 40.0 / 25.0 & 12.0 / 8.0 & 34.0 / 2.0 & 30.0 / 4.0 & 4.0 / 4.0 & 0.0 / 0.0 \\
      \modelicon{qwen_logo.pdf}\hspace{2pt}Qwen3.5-35B
      & & 11.0 / 8.0 & 10.0 / 7.0 & 7.0 / 6.0 & 79.0 / 64.0 & 13.0 / 11.0 & 10.0 / 6.0 & 4.0 / 0.0 & 7.0 / 2.0 & 12.0 / 2.0 & 16.0 / 12.0 \\
      \modelicon{qwen_logo.pdf}\hspace{2pt}Qwen3.5-9B
      & & 17.0 / 10.0 & 18.0 / 11.0 & 18.0 / 18.0 & 92.0 / 72.0 & 18.0 / 13.0 & 10.0 / 4.0 & 6.0 / 0.0 & 8.0 / 0.0 & 8.0 / 4.0 & 16.0 / 16.0 \\

      \midrule
      \rowcolor{gray!8}
      \multicolumn{12}{c}{\textbf{\textit{COM as Action with Retrieval Augmentation}}} \\
      \midrule
      \modelicon{openai_logo.pdf}\hspace{2pt}GPT-5
      & & 30.0 / 6.0 & 50.0 / 36.0 & \underline{70.0} / 52.0 & 49.0 / 40.0 & 18.0 / 2.0 & 45.0 / 31.0 & 0.0 / 0.0 & 14.0 / 2.0 & 48.0 / 28.0 & \underline{65.0} / \underline{43.0} \\
      \modelicon{claude_logo.pdf}\hspace{2pt}Claude-Sonnet-4.6
      & & 10.0 / 4.0 & 68.0 / 58.0 & 49.0 / 40.0 & 25.0 / 25.0 & 11.0 / 8.0 & 36.0 / 30.0 & 0.0 / 0.0 & 0.0 / 0.0 & 40.0 / 25.0 & 9.0 / 7.0 \\
      \modelicon{glm_logo.pdf}\hspace{2pt}GLM-4.6v
      & & 0.0 / 0.0 & 6.0 / 4.0 & 10.0 / 8.0 & 0.0 / 0.0 & 0.0 / 0.0 & 4.0 / 4.0 & 0.0 / 0.0 & 0.0 / 0.0 & 2.0 / 2.0 & 0.0 / 0.0 \\
      \modelicon{qwen_logo.pdf}\hspace{2pt}Qwen3.5-35B
      & & 5.0 / 3.0 & 1.0 / 0.0 & 0.0 / 0.0 & 21.0 / 19.0 & 0.0 / 0.0 & 0.0 / 0.0 & 0.0 / 0.0 & 5.0 / 0.0 & 0.0 / 0.0 & 0.0 / 0.0 \\
      \modelicon{qwen_logo.pdf}\hspace{2pt}Qwen3.5-9B
      & & 0.0 / 0.0 & 1.0 / 1.0 & 4.0 / 2.0 & 0.0 / 0.0 & 0.0 / 0.0 & 0.0 / 0.0 & 0.0 / 0.0 & 0.0 / 0.0 & 0.0 / 0.0 & 2.0 / 0.0 \\

      \midrule

      \rowcolor{slate}
      Ours &
      & \textbf{90.0 / 81.0}
      & \textbf{96.0 / 89.0}
      & \textbf{95.0 / 88.0}
      & 86.0 / \underline{86.0}
      & \textbf{84.0} / \underline{75.0}
      & \textbf{97.0 / 87.0}
      & \textbf{76.0 / 61.0}
      & \textbf{78.0 / 58.0}
      & \textbf{98.0 / 86.0}
      & \textbf{95.0 / 80.0} \\

      \bottomrule
    \end{tabular}%
  }

  \caption{
    Experimental results across single-task and multi-task settings.
    Each cell reports Code Valid Rate / Task Success Rate.
    Abbreviations:
    3D-M = 3D model,
    Asm. = assembly,
    2D-S = 2D sketch,
    E = edit,
    D = drawing,
    MP = mass properties,
    ID = interference detection,
    SW = SolidWorks,
    Inv. = Inventor,
    and ACAD = AutoCAD.
    \textbf{Notably, our method is evaluated without few-shot prompting or retrieval augmentation.}
  }

  \label{tab:experiment-results}
  \vspace{-15pt}
\end{table*}

\section{Experiments}
\vspace{-5pt}
\subsection{Experimental Setup}
\noindent
\textbf{Benchmarks.} To rigorously evaluate agents in authentic engineering workflows, we construct \textbf{ComCADBench}, comprising 1,000 tasks across three mainstream CAD applications: SolidWorks, Inventor, and AutoCAD.
The tasks span seven core activities: 3D modeling, 2D sketching, editing, assembly, engineering drawing, mass property analysis, and interference detection.
The benchmark is structured into two levels: Single-Task containing 400 samples for fundamental operations and Multi-Task containing 600 samples for long-horizon pipelines such as interference detection after assembly.
Tasks are sourced from three public datasets Text2CAD~\cite{ref:text2cad}, Fusion360 Gallery~\cite{ref:fusion360gallery}, SketchGraphs~\cite{ref:sketchgraphs}, and supplemented with human-crafted instructions to cover real engineering workflows.
Each sample includes a text instruction and a human-verified Ground Truth (GT) output, rigorously validated within our ComForge environment to ensure executability.
A brief statistics table is provided in Table~\ref{tab:comcadbench_statistics}.
\noindent We additionally evaluate on two public CAD generation datasets \textbf{Text2CAD}~\cite{ref:text2cad} and \textbf{CADPrompt}~\cite{ref:cadprompt}, containing 8,042 and 200 test cases respectively.

\noindent \textbf{Evaluation Metrics.}
We report two primary metrics: \textbf{Valid Rate (VR)} measures the percentage of error-free executing COM code, and \textbf{Success Rate (SR)} evaluates whether final artifacts pass task-specific criteria.
Specifically,
(1) 3D modeling \& assembly success requires an STL-based Chamfer Distance of $\mathrm{CD} \le 10^{-3}$ against the ground truth (GT);
(2) 2D sketches are evaluated via DXF-based $\mathrm{CD}$ with a strict threshold of $10^{-30}$;
(3) engineering drawings must contain all required views and annotations; and
(4) mass properties \& interference require an element-wise JSON comparison against the GT. Multi-task pipelines strictly require passing all constituent sub-tasks to achieve overall success.


\noindent \textbf{Baselines and Implementation Details.}
To evaluate our paradigm, we select three categories of baselines:
(1) GUI-based agents (GPT-5~\cite{ref:gpt-5}, Qwen3.5-9B~\cite{ref:qwen3.5}) operating via screen coordinates and pixel actions;
(2) COM-based general LLMs (GPT-5~\cite{ref:gpt-5}, Claude-Sonnet-4.6~\cite{ref:claude-sonnet-4-6}, GLM-4.6v~\cite{ref:glm-4.6v}, Qwen3.5-35B/9B~\cite{ref:qwen3.5}) tested under zero-shot, few-shot, and RAG settings; and
(3) Domain-Specific CAD models (Text2CAD~\cite{ref:text2cad}, CAD-Coder~\cite{ref:cadcoder}, CADmium~\cite{ref:cadprompt}, CAD-Judge~\cite{zhou2026cad}).
For implementation, \actor{} is initialized from Qwen3.5-9B and trained using the \texttt{ms-swift} framework on 8 A100 GPUs within ComForge. The training pipeline utilizes 12k samples for Stage 1, 30k for Stage 2, and 2,156 samples for Stage 3 (GRPO) focused on 3D modeling, assembly, and 2D sketches. During evaluation, GUI and COM agents are limited to 35 interactive turns and 5 iterative code modifications, respectively. Details are provided in Appendix.

\subsection{Main Results}
\vspace{-5pt}

\noindent \textbf{SOTA Performance on \bench{}.}
As shown in Table~\ref{tab:experiment-results}, our 9B \actor{} establishes a new state-of-the-art across all categories, outperforming proprietary giants like GPT-5 and Claude-Sonnet-4.6 in both single- and multi-task setups. This advantage peaks in multi-task pipelines, where early-stage geometric flaws typically trigger catastrophic failures downstream. While top proprietary baselines suffer severe performance degradation under these tight task dependencies, \actor{} consistently maintains an end-to-end success rate above 80\%. This striking resilience proves that our progressive training framework does more than teach API syntax—it successfully empowers the agent to parse environment feedback, detect latent geometric mismatches, and dynamically self-correct its code before errors can cascade.

\begin{table}[t]
  \centering
  \renewcommand{\arraystretch}{1.12}
  \newcolumntype{Y}{>{\centering\arraybackslash}X}

  \begin{tabularx}{0.8\linewidth}{lYYY}
    \toprule

    \textbf{Method} & \textbf{Mean CD $\downarrow$} & \textbf{Median CD $\downarrow$} & \textbf{IR $\downarrow$} \\

    \midrule

    Claude-3.7-Sonnet & 186.53 & 134.16 & 47.03 \\
    GPT-4o            & 133.52 & 45.91  & 93.00 \\
    DeepSeek-V3       & 186.69 & 107.57 & 51.96 \\
    Qwen2.5-72B       & 209.41 & 153.81 & 82.64 \\
    Qwen2.5-7B        & 202.35 & 169.86 & 98.83 \\
    Text2CAD          & 29.29  & 0.37   & 3.75 \\
    CAD-Coder         & \underline{6.54} & 0.17 & \textbf{1.45} \\

    \midrule
    \rowcolor{slate}
    Ours-SolidWorks   & 7.63 & \underline{0.09} & 10.86 \\
    \rowcolor{slate}
    Ours-Inventor     & \textbf{1.20} & \textbf{0.07} & \underline{3.25} \\

    \bottomrule
  \end{tabularx}

  \vspace{10pt}
  \caption{
    Text2CAD full-set evaluation results.
    CD denotes Chamfer Distance and IR denotes invalid ratio. CD values are multiplied by $10^3$.
  }

  \label{tab:generalization-text2cad}
\end{table}

\begin{table}[t]
  \centering
  \renewcommand{\arraystretch}{1.12}

  \begin{tabularx}{0.6\linewidth}{Xccc}
    \toprule

    \textbf{Method} & \textbf{Mean CD $\downarrow$} & \textbf{Median CD $\downarrow$} & \textbf{IR $\downarrow$} \\

    \midrule

    Text2CAD  & --     & 127.70 & \textbf{1.57} \\
    CADmium   & --     & 116.75 & 6.28 \\
    CAD-Judge & 152.40 & 42.69  & \underline{2.55} \\

    \midrule
    \rowcolor{slate}
    Ours-SolidWorks & \textbf{24.15} & \textbf{6.07} & 11.25 \\
    \rowcolor{slate}
    Ours-Inventor   & \underline{26.47} & \underline{6.69} & 7.5 \\

    \bottomrule
  \end{tabularx}

  \vspace{5pt}
  \caption{
    CADPrompt benchmark results.
  }

  \label{tab:generalization-cadprompt}
\end{table}

\noindent \textbf{Evaluation on Text2CAD and CADPrompt.}
We further evaluate \actor{} on two standard CAD generation benchmarks, Text2CAD and CADPrompt, with results reported in Table~\ref{tab:generalization-text2cad} and Table~\ref{tab:generalization-cadprompt}. Both benchmarks report CD to measure geometric deviation between the generated 3D shape and the ground truth, alongside the IR (=1-VR) to reflect the proportion of output code that fails to execute.

On Text2CAD, shown in Table~\ref{tab:generalization-text2cad}, \actor{} with the Inventor backend achieves the best CD metrics (Mean 1.20, Median 0.07), surpassing the specialized model CAD-Coder (6.54 / 0.17) by a clear margin. This is achieved despite training on only the 7k official split rather than the full dataset. While our IR is slightly higher (3.25\% vs. 1.45\%), \actor{} delivers superior geometric precision when the code executes successfully.

On the out-of-domain CADPrompt dataset, \actor{} is tested in a zero-shot setting and establishes a dominant advantage in geometric fidelity, as shown in Table~\ref{tab:generalization-cadprompt}. With the SolidWorks backend, it achieves a Mean CD of 24.15 and a Median CD of 6.07, outperforming CADmium (116.75 / 6.28) and Text2CAD (127.7) by an order of magnitude. Although the strict syntactic complexity of COM code leads to a relatively higher IR on unseen shape types (7.50\% vs. 1.57\%), the successfully compiled outputs are substantially more accurate. These consistent advantages across both benchmarks confirm the strong generalization and robust geometric reasoning of \actor{}.

\begin{figure*}[!t]
  \centering
  \includegraphics[width=\linewidth]{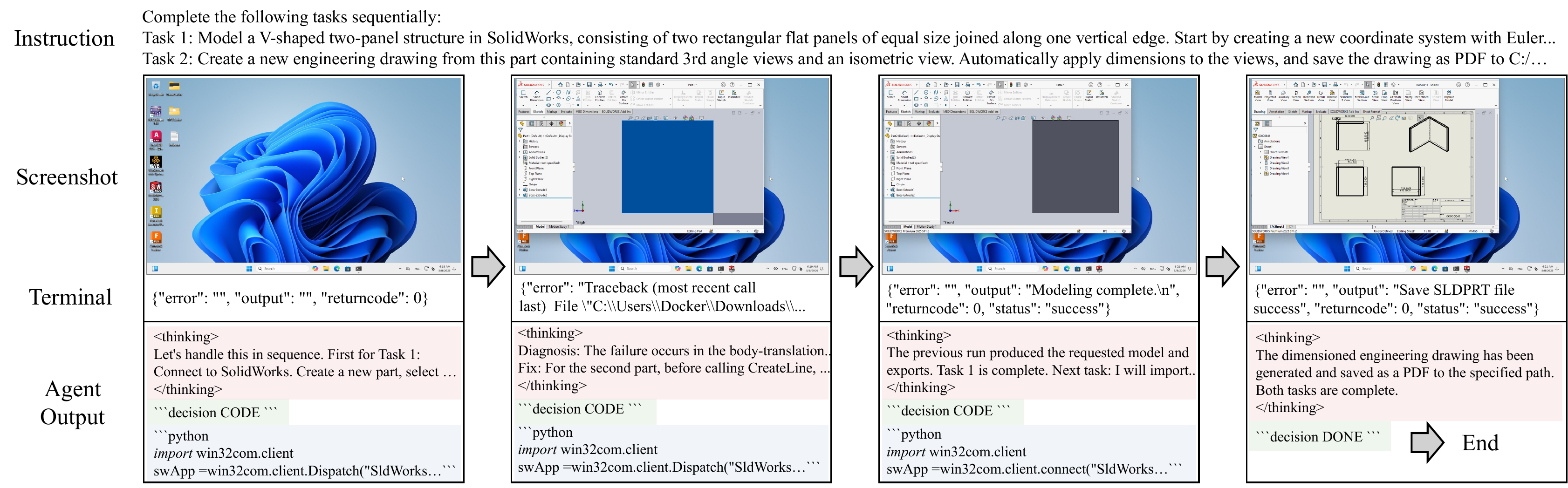}
  \caption{
    An execution trajectory of our agent completing a multi-task pipeline (modeling and engineering drawing).
  }

  \label{fig:traj}
  \vspace{-10pt}
\end{figure*}

\subsection{Insights into CAD Interaction Paradigms}

We analyze baseline behaviors across interaction paradigms (Table~\ref{tab:experiment-results}) and highlight three key insights; detailed analysis is provided in Appendix.
\noindent\textbf{1. GUI vs. COM Paradigm Shift.} GUI-based execution universally collapses (0.0\% SR) due to visual grounding errors and error accumulation over long horizons. Shifting to COM instantly unlocks non-trivial zero-shot performance (e.g., GPT-5 reaches 33.0\% SR on 2D sketches), confirming that LLMs can project general coding proficiency into CAD via programmatic backends.
\noindent\textbf{2. Syntax-Geometry Gap.} Few-shot prompting improves the Valid Rate, yet a severe gap persists between syntactic correctness and geometric precision. Baselines frequently synthesize error-free COM code that violates engineering constraints, highlighting the necessity of \actor{}'s RL stage, which directly optimizes geometric intent rather than syntactic form.
\noindent\textbf{3. Negative Effects of RAG.} RAG counter-intuitively degrades performance (e.g., GPT-5's SR drops on 3D modeling), as retrieved API patterns introduce idiomatic inconsistencies and RAG prompts displace critical task context in multi-task pipelines.

\subsection{Ablation Study}

We ablate the three training stages of \actor{} on the single-task subset, with results summarized in Table~\ref{tab:ablation}. Each progressive stage yields a clear performance gain, validating the overall effectiveness and necessity of our multi-stage training framework.
Stage 1 provides a foundational baseline for code generation, achieving valid/success rates of 58.0\% / 52.0\% on SW, 89.0\% / 79.0\% on Inv., and 73.0\% / 66.0\% on Asm. Introducing multi-turn interaction in Stage 2 delivers the largest gain in Code Valid Rate (boosting SW to 88.0\%, Inv. to 95.0\%, and Asm. to 93.0\%), as it enables the agent to dynamically recover from initial execution errors using environment feedback.
However, after Stage 2, a persistent gap between code validity and task success remains (e.g., a 12.0\% gap on SW and an 11.0\% gap on Asm.). This indicates that the agent occasionally settles once the code compiles, overlooking underlying geometric mistakes. Stage-3 Reinforcement Learning (RL) explicitly closes this gap by optimizing task-level rewards on a curated set of high-quality data. It boosts the final performance to \textbf{90.0\% / 81.0\%} on SW, \textbf{96.0\% / 89.0\%} on Inv., and \textbf{95.0\% / 88.0\%} on Asm., strictly narrowing the discrepancy between code validity and task completion, and successfully aligning the agent's programmatic behavior with the user's true geometric intent.













\newcolumntype{L}{>{\raggedright\arraybackslash}X}

\begin{table}[t]
  \centering
  \renewcommand{\arraystretch}{1.12}

  \begin{tabularx}{0.7\linewidth}{Lccc}
    \toprule

    \textbf{Method} & \multicolumn{2}{c}{\textbf{3D-M}} & \textbf{Asm.} \\
    \cmidrule(lr){2-3} \cmidrule(lr){4-4}
    \textbf{VR / SR}\,$\uparrow$ & \textbf{SW} & \textbf{Inv.} & \textbf{Inv.} \\

    \midrule

    Ours + SFT-stage1
    & 58.0 / 52.0 & 89.0 / 79.0 & 73.0 / 66.0 \\

    Ours + SFT-stage1\&2
    & 88.0 / 76.0 & 95.0 / 87.0 & 93.0 / 82.0 \\

    Ours + SFT + RL
    & \textbf{90.0 / 81.0}
    & \textbf{96.0 / 89.0}
    & \textbf{95.0 / 88.0} \\

    \bottomrule
  \end{tabularx}

  \vspace{5pt}
  \caption{Ablation study across different training stages.}

  \label{tab:ablation}

  \vspace{-15pt}
\end{table}

\section{Conclusion}
We introduce \method{}, a new interaction paradigm that reframes professional software manipulation as COM program synthesis, establishing a unified action space across diverse industrial applications. To operationalize this paradigm, we construct \bench{}, the first benchmark grounded in real CAD software, build ComForge, a mass-parallelized infrastructure for real-environment training and evaluation, and develop a self-correcting agent \actor{} via a progressive three-stage training framework.
 Extensive experiments show that \actor{} achieves superior performance on \bench{} and generalizes well to Text2CAD and CADPrompt. 
 We hope 
 our paradigm, benchmark, and infrastructure 
 can serve as a foundation for future research on agentic automation in professional software environments.
\section*{Limitations}
Despite the promising results of COM-as-Action, our current framework still has several limitations. Due to licensing constraints of commercial professional software, ComForge currently provides non-activated software environments, while evaluation is conducted through a centralized submission platform. In addition, our experiments are currently limited to CAD software and the Windows ecosystem, primarily because many widely used professional applications expose native COM interfaces only on Windows. Nevertheless, COM-as-Action is not inherently restricted to CAD and could potentially generalize to broader professional software ecosystems such as Office, Adobe, and other engineering applications. Finally, ComActor currently relies on relatively limited historical context during iterative code refinement. Future work may explore stronger memory and planning mechanisms, such as structured execution history or long-horizon memory, to improve reasoning and self-correction in complex workflows.

\clearpage
{
  \bibliographystyle{unsrt}  
  \bibliography{preprint}
}

\clearpage
\newgeometry{
  textheight=9in, textwidth=5.5in, top=1in,
  headheight=12pt, headsep=25pt, footskip=30pt
}

\appendix
\cleardoublepage
\appendix
\noindent{\LARGE\textbf{Appendix}\par}\normalsize

\makeatletter
\newcommand{\appendixtableofcontents}{%
  \begingroup
  \section*{Contents}%
  \@starttoc{atoc}%
  \endgroup
}
\appendixtableofcontents
\renewcommand{\addcontentsline}[3]{%
  \begingroup
  \def\app@type{#2}%
  \def\app@section{section}%
  \def\app@subsection{subsection}%
  \ifx\app@type\app@section
  \addtocontents{atoc}{\protect\contentsline{#2}{#3}{\thepage}{}}%
  \else\ifx\app@type\app@subsection
  \addtocontents{atoc}{\protect\contentsline{#2}{#3}{\thepage}{}}%
  \fi\fi
  \endgroup
}
\makeatother
\vspace{2mm}

\section{Data Construction}
To ensure the high quality and practical relevance of ComCADBench, we employ a rigorous, multi-stage data construction pipeline. This process transforms raw parametric data into natural language instructions paired with executable COM scripts, verified through physical execution in our ComForge environment.
\subsection{Source Data and Textualization}
We construct the foundation of our benchmark by sourcing high-quality parametric CAD designs from three public datasets: Text2CAD~\cite{ref:text2cad} (for 3D modeling), SketchGraphs~\cite{ref:sketchgraphs} (for 2D sketching), and Fusion360Gallery~\cite{ref:fusion360gallery} (for assembly). These platforms natively provide structured parametric descriptions of models in JSON format, which contain all necessary geometric entities and topological constraints.

Since SketchGraphs and Fusion360Gallery do not provide natural language instructions, we develop an automated textualization pipeline. We prompt a frontier MLLM (GPT-5~\cite{ref:gpt-5}) to ingest the structured JSON files and generate corresponding professional engineering instructions. To guarantee linguistic diversity and professional accuracy, a subset of the generated text instructions undergoes rigorous human evaluation. For Text2CAD, we directly utilize its high-quality human-annotated text instructions.

\subsection{Ground Truth COM Script Synthesis}
With the structured JSON files containing complete topological and geometric information, we employ a rule-based engine to synthesize the Ground Truth (GT) COM scripts. For instance, in 2D sketch tasks, our parser extracts entity types (e.g., lines, arcs, circles) and their precise coordinates from the JSON, subsequently mapping them to native COM APIs to generate the deterministic executable script.
To illustrate this mapping, Table~\ref{tab:json_to_com} provides a side-by-side comparison of the raw JSON representation and the synthesized COM script for a standard sketching operation.
\begin{table*}[h]
\centering
\small
\begin{minipage}[t]{0.48\linewidth}
\centering
\textbf{Parametric JSON Representation}
\vspace{0.5em}
\begin{verbatim}
{
    "sketch_id": "f77bdbb84580a54d22076c0e_0_0",
    "entities": [
        {
            "type": "Line",
            "entityId": "1",
            "isConstruction": false,
            "dirX": 1.0,
            "dirY": 0.0,
            "pntX": 0.0014664754271507263,
            "pntY": 0.012719333171844482,
            "startParam": -0.07332348078489304,
            "endParam": 0.07332348078489304
        },
        {
            "type": "Line",
            "entityId": "4",
            "isConstruction": false,
            "dirX": 1.0,
            "dirY": 0.0,
            "pntX": 0.0014664754271507263,
            "pntY": 0.0,
            "startParam": -0.07332348078489304,
            "endParam": 0.07332348078489304
        },
        {
            "type": "Line",
            "entityId": "7",
            "isConstruction": false,
            "dirX": 0.0,
            "dirY": -1.0,
            "pntX": -0.07185700535774231,
            "pntY": 0.006359666585922241,
            "startParam": -0.006359666585922241,
            "endParam": 0.006359666585922241
        },
        {
            "type": "Line",
            "entityId": "10",
            "isConstruction": false,
            "dirX": 0.0,
            "dirY": -1.0,
            "pntX": 0.07478995621204376,
            "pntY": 0.006359666585922241,
            "startParam": -0.006359666585922241,
            "endParam": 0.006359666585922241
        },
        {
            "type": "Arc",
            "entityId": "13",
            "isConstruction": false,
            "xCenter": -0.0718570053577423,
            "yCenter": 0.006359666585922241,
            "xDir": 1.0,
            "yDir": 0.0,
            "radius": 0.006359666585922241,
            "clockwise": false,
            "startParam": -4.712388980384688,
            "endParam": -1.5707963267948988
        }
    ]    
}
\end{verbatim}
\end{minipage}
\hfill
\begin{minipage}[t]{0.48\linewidth}
\centering
\textbf{Synthesized GT COM Script (Python)}
\vspace{0.5em}
\begin{verbatim}
# --- Sketch ---
# Entity 1
points_1 = get_line_points(
    pnt_x = 0.0014664754271507263,
    pnt_y = 0.012719333171844482,
    dir_x = 1.0,
    dir_y = 0.0,
    start_param = -0.07332348078489304,
    end_param = 0.07332348078489304,
)
start_pnt_1 = win32com.client.VARIANT((
    points_1['start_x']*M_to_MM, 
    points_1['start_y']*M_to_MM, 
    0.0
))
end_pnt_1 = win32com.client.VARIANT((
    points_1['end_x']*M_to_MM, 
    points_1['end_y']*M_to_MM, 
    0.0
))
model_space.AddLine(start_pnt_1, end_pnt_1)

# Entity 4
points_4 = get_line_points(
    pnt_x = 0.0014664754271507263,
    pnt_y = 0.0,
    dir_x = 1.0,
    dir_y = 0.0,
    start_param = -0.07332348078489304,
    end_param = 0.07332348078489304,
)
start_pnt_4 = win32com.client.VARIANT((
    points_4['start_x']*M_to_MM, 
    points_4['start_y']*M_to_MM, 
    0.0
))
end_pnt_4 = win32com.client.VARIANT((
    points_4['end_x']*M_to_MM, 
    points_4['end_y']*M_to_MM, 
    0.0
))
model_space.AddLine(start_pnt_4, end_pnt_4)

# Entity 7
points_7 = get_line_points(
    pnt_x = -0.07185700535774231,
    pnt_y = 0.006359666585922241,
    dir_x = 0.0,
    dir_y = -1.0,
    start_param = -0.006359666585922241,
    end_param = 0.006359666585922241,
)
start_pnt_7 = win32com.client.VARIANT((
    points_7['start_x']*M_to_MM, 
    points_7['start_y']*M_to_MM, 
    0.0
))
end_pnt_7 = win32com.client.VARIANT((
    points_7['end_x']*M_to_MM, 
    points_7['end_y']*M_to_MM, 
    0.0
))
model_space.AddLine(start_pnt_7, end_pnt_7)
...
\end{verbatim}
\end{minipage}
\caption{An example of mapping raw parametric JSON data to the corresponding COM script.}
\label{tab:json_to_com}
\end{table*}

\begin{table*}[tbp]
  \centering
  \small
  \setlength{\tabcolsep}{8pt} 
  \begin{tabular}{@{}lllr@{}}
    \toprule
    \textbf{Complexity} & \textbf{Task Category} & \textbf{Software} & \textbf{\# Samples} \\
    \midrule
    \multirow{3}{*}{\shortstack[l]{\textbf{Single-Task}\\($N=400$)}}
    & 3D Modeling & SolidWorks, Inventor & 200  \\
    & Assembly & Inventor & 100  \\
    & 2D Draft / Sketch & AutoCAD & 100  \\
    \midrule
    \multirow{4}{*}{\shortstack[l]{\textbf{Multi-Task}\\($N=600$)}}
    & 3D Modeling + Edit & SolidWorks, Inventor & 200  \\
    & 3D Modeling + Drawing & SolidWorks, Inventor & 100  \\
    & 3D Modeling + Mass Prop. & SolidWorks, Inventor & 200  \\
    & Assembly + Interference & Inventor & 100  \\
    \midrule
    \textbf{Total} & \textbf{All 7 Categories} & \textbf{SW, Inv, ACAD} & \textbf{1,000} \\
    \bottomrule
  \end{tabular}
  \caption{Overall statistics of \textbf{ComCADBench}.}
  \label{tab:comcadbench_statistics}
\end{table*}

\begin{table*}[h]
\centering
\small
\begin{tabular}{lcc}
\toprule
\textbf{CAD Software} & \textbf{Total Crawled} & \textbf{Task-Related Filtered} \\
\midrule
SolidWorks & 12,376 (Interfaces) + 961 (Const) & 3,302 + 38 \\
Inventor & 24,269 (Objects) + 451 (Enums) & 1,509 + 11 \\
AutoCAD & 4,078 (Drawing/File only) & 4,078 \\
\bottomrule
\end{tabular}
\caption{Statistics of the crawled COM API database across three CAD platforms. The filtered subset represents APIs directly relevant to the core engineering tasks evaluated in \bench{}.}
\label{tab:api_crawl_stats}
\end{table*}
\begin{table}[h]
\centering
\small
\begin{tabular}{lc}
\toprule
\textbf{Task Category \& Software} & \textbf{\makecell[c]{Retrieved APIs \\ (Avg / Max / Min)}} \\
\midrule
\textbf{2D Sketch} & \\
- AutoCAD & 9.45 / 10 / 7 \\
\midrule
\textbf{3D Model} & \\
- SolidWorks & 17.45 / 24 / 12 \\
- Inventor & 38.17 / 39 / 35 \\
\midrule
\textbf{Assembly} & \\
- Inventor & 26.00 / 26 / 26 \\
\midrule
\textbf{3D Model + Modify} & \\
- SolidWorks & 17.98 / 20 / 12 \\
- Inventor & 38.30 / 39 / 35 \\
\midrule
\textbf{3D Model + Draw} & \\
- SolidWorks & 30.45 / 37 / 25 \\
\midrule
\textbf{3D Model + Mass Property} & \\
- SolidWorks & 33.06 / 38 / 27 \\
- Inventor & 47.40 / 49 / 41 \\
\midrule
\textbf{Assembly + Interference} & \\
- Inventor & 35.00 / 35 / 35 \\
\bottomrule
\end{tabular}
\caption{Number of retrieved APIs for tasks across task categories and software.}
\label{tab:task_retrieval_stats}
\end{table}

\subsection{Downstream Multi-Task Construction}
Beyond foundational modeling, real engineering workflows require diverse downstream operations. We construct the remaining task categories as follows:

\begin{itemize}
  \item \textbf{Mass Properties Analysis:} We predefine a comprehensive list of analytical properties, including mass, volume, density, center of mass, principal moments, principal rotation (for Inventor), principal axes (for SolidWorks), inertia tensor at centroid, and inertia tensor at origin. We randomly pair these properties with 8 standard industrial materials (Water, Carbon Steel, Alloy Steel, Stainless Steel 304, Aluminum 6061, Brass, ABS Plastic, Nylon 6/6) to generate instructions and construct the corresponding GT scripts.
  \item \textbf{Interference Detection:} We implement two levels of collision analysis: Hard Interference, which checks for physical volume overlaps between intersecting components, and Soft Interference, which evaluates clearance boundaries and bounding box intersections. GT scripts are generated to sequentially invoke the respective CAD detection APIs.
  \item \textbf{Model Editing:} We define a set of fundamental modification operations: adding/deleting entities, modifying dimensions, applying fillets/chamfers, and drilling holes. We prompt GPT-5 to select from these operations, formulate an edit instruction, and modify the original modeling COM script accordingly.
\end{itemize}

\noindent
\textbf{ Cross-Model Consensus for Edit Verification:} Since LLM-generated edit scripts cannot be guaranteed correct by rule-based generation alone, we introduce a dual-model consensus mechanism. Using the same edit instruction and original script, we prompt a secondary model (Qwen-32B) to generate an independent edit script. Both scripts are executed, and if the Chamfer Distance (CD) between their resulting 3D artifacts is below a strict threshold ($\epsilon \le 10^{-4}$), the sample is deemed valid. We retain the edit instruction, the GT code, and the output artifact. A random subset is additionally verified by human experts to ensure dataset reliability.

\noindent
\textbf{Multi-Task Pipelines:} Long-horizon tasks are constructed by composing single-turn instructions sequentially (e.g., executing a 3D assembly followed immediately by interference detection).

\subsection{Execution Verification in ComForge}
To guarantee strict executability, all synthesized GT scripts across all task categories are deployed and executed within the Dockerized Windows environments of ComForge. The artifacts generated by the scripts are extracted and compared against the original dataset's ground truth models. Only samples whose script-generated artifacts achieve a Chamfer Distance below the strict threshold against the true GT are retained in the final ComCADBench dataset.

\subsection{Data Visualization and Statistics}
\begin{figure*}[t]
  \centering
  \includegraphics[width=\linewidth]{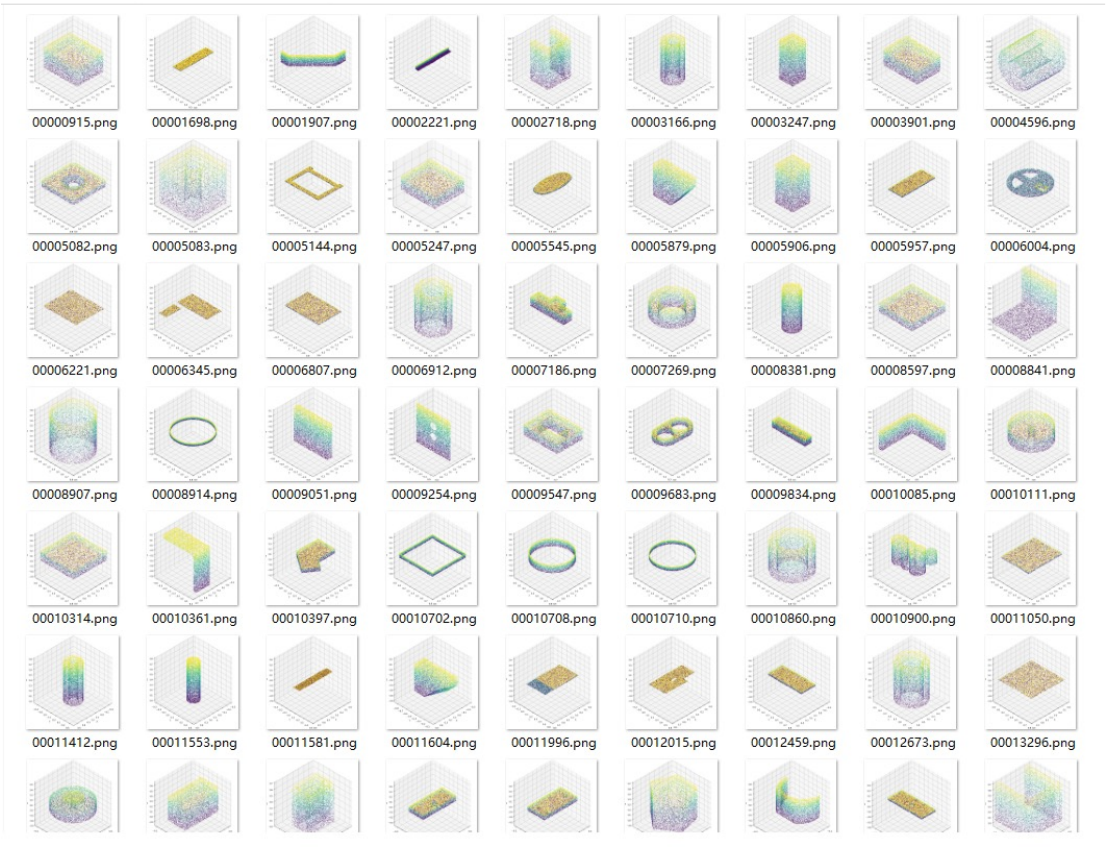}
  \caption{Visualization of the ground truth artifacts for 3d modeling samples in \bench{}.}
  \label{fig:sample_text2cad}
\end{figure*}

\begin{figure*}[t]
  \centering
  \includegraphics[width=\linewidth]{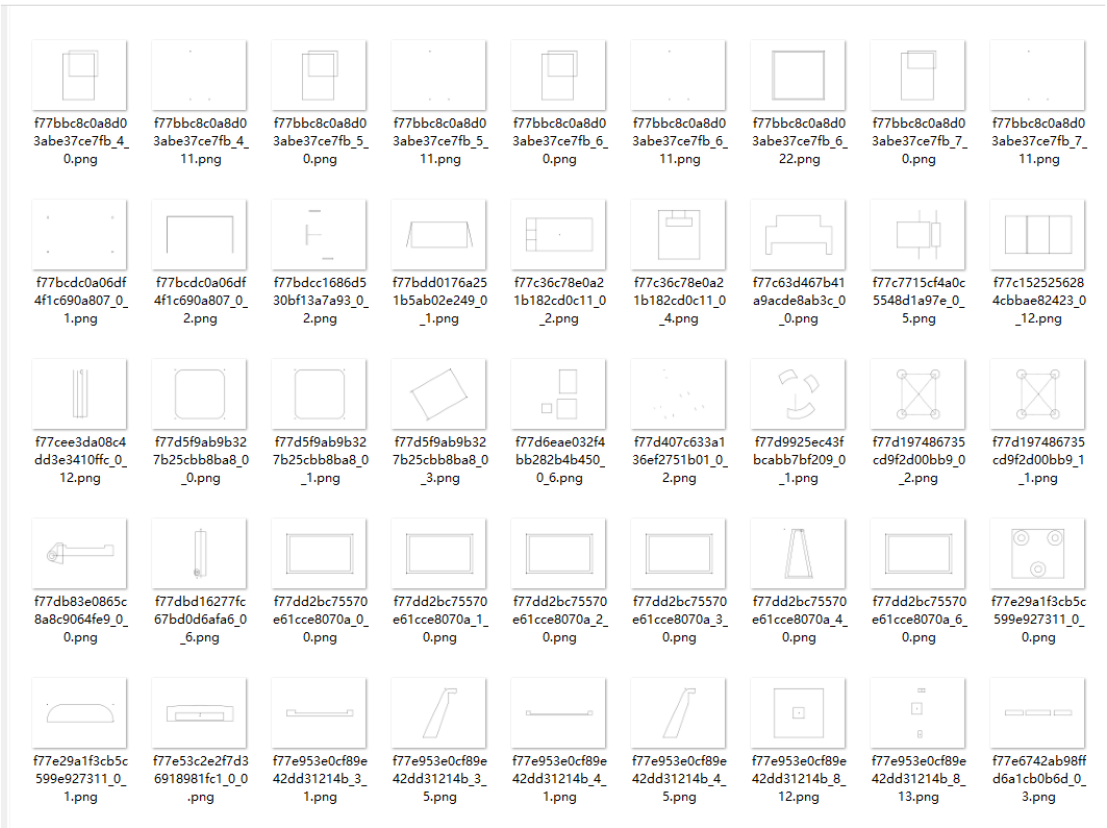}
  \caption{Visualization of the ground truth artifacts for 2d sketching samples in \bench{}.}
  \label{fig:sample_sketchgraphs}
\end{figure*}

\begin{figure*}[t]
  \centering
  \includegraphics[width=\linewidth]{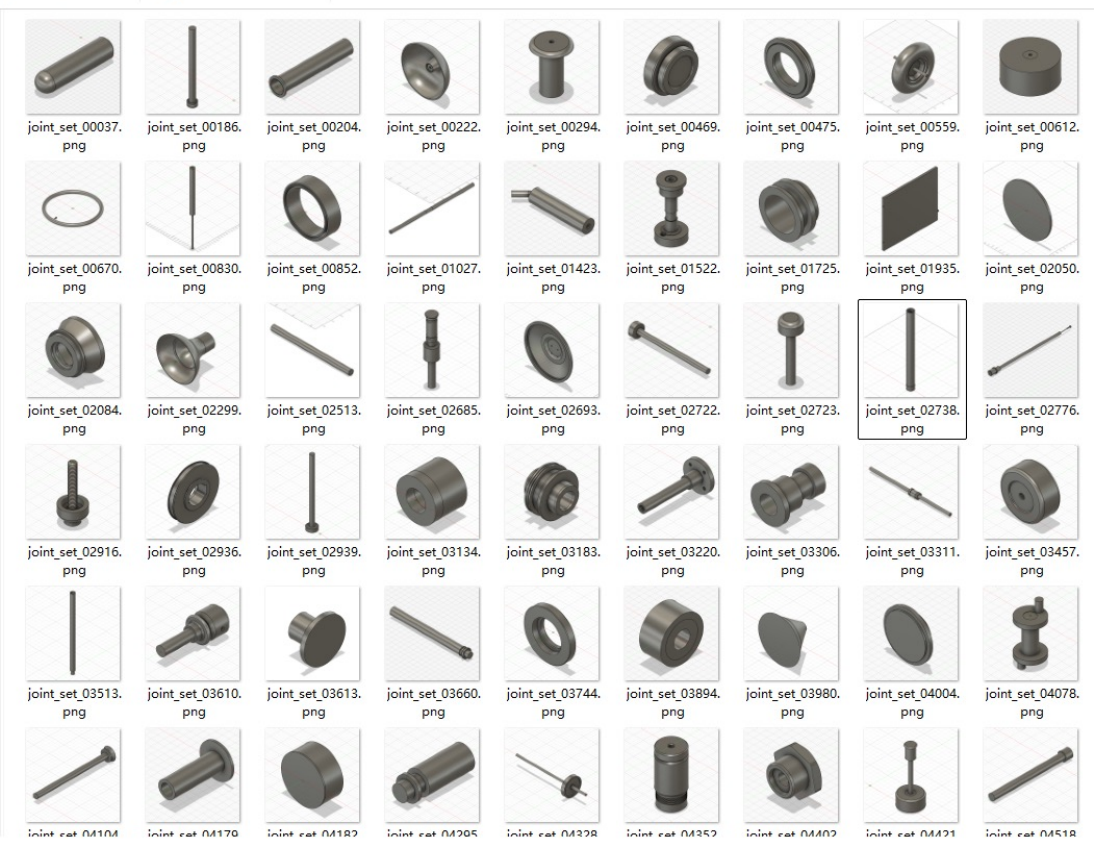}
  \caption{Visualization of the ground truth artifacts for assembly samples in \bench{}.}
  \label{fig:sample_assembly}
\end{figure*}

\begin{figure*}[t]
  \centering
  \includegraphics[width=\linewidth]{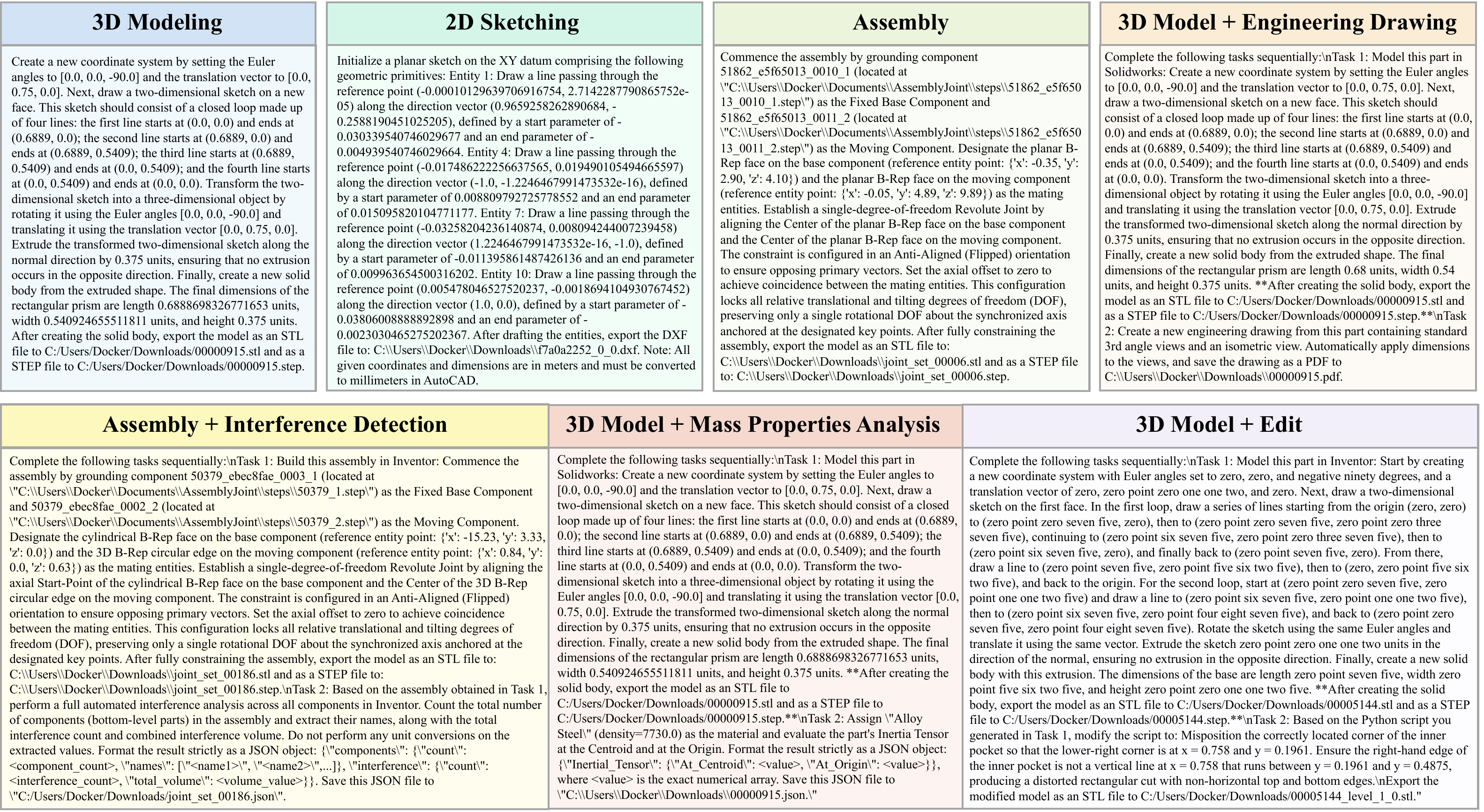}
  \caption{Detailed examples of input instructions across all specific task categories in \bench{}. The examples highlight the professional engineering terminology, precise geometric constraints, and complex workflows required to complete each automated CAD operation.}
  \label{fig:task_details}
\end{figure*}

The detailed statistical distribution of ComCADBench across different task types and software platforms is presented in Table~\ref{tab:comcadbench_statistics}. We visualize GT artifacts in representative samples for 3D modeling (Figure~\ref{fig:sample_text2cad}), assembly (Figure ~\ref{fig:sample_assembly}), and 2D sketching (Figure~\ref{fig:sample_sketchgraphs}) tasks. Figure~\ref{fig:task_details} provides instruction examples for each specific task category.


\subsection{Detailed Empirical Insights into CAD Paradigms}

To better understand the mechanisms driving CAD automation, we analyze the performance of various baseline models under different interaction paradigms and prompt configurations. This analysis reveals several critical insights:

\noindent \textbf{Limitations of GUI-Based Paradigms.} GUI-based paradigms fundamentally fail on professional CAD tasks. As evidenced by the top rows of Table~\ref{tab:experiment-results}, both GPT-5 and Qwen3.5-9B achieve a 0.0\% SR and VR across all categories. Industrial CAD interfaces feature dense micro-icons and complex 3D viewports where MLLMs struggle with precise pixel-level grounding. Furthermore, continuous spatial reasoning errors accumulate inexorably across long-horizon trajectories, causing the execution to collapse before task completion.

\noindent \textbf{Unlocking Full Reasoning Capability via the COM Paradigm.} By shifting the interaction from fragile visual perception to deterministic code synthesis, the COM paradigm instantly unlocks non-trivial performance for proprietary models without domain-specific training. Table~\ref{tab:experiment-results} clearly highlights this step-function improvement: in a zero-shot setting, GPT-5 and Claude-Sonnet-4.6 immediately jump from complete failure to meaningful task completion, with GPT-5 achieving up to 71.0\% VR and 33.0\% SR on 2D sketches. This validates that generalist LLMs can project their inherent coding proficiency into professional software via COM.

\noindent \textbf{Task Complexity and Scale Discrepancies.} Tasks with well-structured, flat API surfaces benefit most from the COM paradigm. 2D sketching in AutoCAD, for instance, maps cleanly onto a linear sequence of coordinate-based API calls, enabling GPT-5 to reach 58\% SR with few-shot prompting. 3D modeling, in contrast, demands multi-coordinate-system reasoning, feature tree manipulation, and geometric constraint solving, which remains challenging even for strong models. Furthermore, a pronounced gap separates proprietary models from open-source alternatives. Qwen3.5-9B and Qwen3.5-35B barely exceed zero on most tasks without in-context support, indicating that effective COM manipulation relies heavily on pre-training exposure to domain-specific APIs. This gap widens further on multi-task pipelines, where errors in early sub-tasks propagate across stages, causing all untrained baselines to degrade sharply.

\noindent \textbf{The Syntax-Geometry Gap in Advanced Prompting.} While few-shot prompting significantly elevates the code valid rate, a massive discrepancy remains between code validity and geometric accuracy for generalist LLMs. This reveals a fundamental limitation: models can successfully mimic COM syntax to avoid execution crashes, yet lack the spatial reasoning required to satisfy engineering constraints. They frequently produce code that runs perfectly but yields geometrically incorrect results. This gap highlights the necessity of our RL alignment, which explicitly targets geometric intent rather than syntactic correctness.

\noindent \textbf{Counterproductive Effects of Retrieval-Augmented Generation.} On several tasks, RAG even degrades performance relative to zero‑shot prompting. This is particularly noticeable for GPT‑5, where additional API descriptions appear to disrupt rather than complement the model's existing COM knowledge. We attribute this to two factors: first, the retrieval corpus cannot distinguish between common and rare API usage patterns, occasionally surfacing functionally correct but idiomatically awkward APIs that lead to syntactically valid yet geometrically nonsensical code. Second, the RAG prompt occupies a substantial portion of the context window, displacing critical information about task constraints and environment states. This negative effect is most pronounced on multi‑task pipelines, where coordinating across sub‑tasks already strains the model's context utilization.

\section{Implementation Details}
\subsection{Training Implementation and Hyperparameters}
All training stages of ComActor are conducted on 8 NVIDIA A100 (80GB) GPUs. Our model is initialized from the Qwen3.5-9B~\cite{ref:qwen3.5} backbone. The training pipeline is implemented using the ms-swift framework, which provides robust support for both parameter-efficient fine-tuning and large-scale Reinforcement Learning from Human Feedback (RLHF).

\noindent
\textbf{Stage 1 \& Stage 2 (Supervised Fine-Tuning):} For both single-turn (Stage 1) and multi-turn (Stage 2) SFT, we apply Low-Rank Adaptation (LoRA) to all linear layers with a rank of $r=8$ and $\alpha=32$. The models are trained using the AdamW optimizer with a learning rate of 1e-5, a cosine learning rate scheduler, and a warmup ratio of 0.05. To accommodate the extensive context required for code generation and error tracebacks, the maximum sequence length is set to 16,384 tokens. DeepSpeed ZeRO-2 is enabled to optimize memory consumption.

\noindent
\textbf{Stage 3 (Group Relative Policy Optimization - GRPO):} For the RL stage, we transition to full-parameter fine-tuning to maximize the model's capacity for geometric alignment. The training utilizes 2,156 curated samples focusing on 3D modeling, assembly, and 2D sketching. We integrate vLLM directly into the ms-swift pipeline (Colocate mode) to accelerate policy rollout generation. For each prompt, the model samples 4 distinct completions (generations) with a temperature of 1.0. To evaluate the rewards, we deploy 64 concurrent Dockerized Windows VMs within our ComForge infrastructure, providing synchronous real-time feedback. DeepSpeed ZeRO-3 is employed for distributed training.

\subsection{Baseline Configurations and Prompting Strategies}
To systematically evaluate the capability of generalist LLMs on CAD manipulation, we test baselines under Zero-shot, Few-shot, and Retrieval-Augmented Generation (RAG) settings. The system prompt universally assigns the model a professional role, specifies the target software, and strictly constrains the output format to executable Python COM scripts enclosed in markdown code blocks. Detailed prompts under each setting are provided in Appendix~\ref{sec:prompts}.

\noindent
\textbf{Few-Shot Prompting.}
For complex spatial tasks such as 3D modeling and multi-task pipelines (assembly followed by interference detection), pure zero-shot generation often struggles with precise coordinate transformations and topological constraints. In the few-shot setting, we prepend one high-quality demonstration exemplar to the prompt context of each task category. These exemplars are manually curated to demonstrate the standard API workflow and correct spatial reasoning patterns for the target software.

\noindent
\textbf{Retrieval-Augmented Generation.}
To equip generalist MLLMs with the necessary domain knowledge for COM programming, we implement a Retrieval-Augmented Generation (RAG) pipeline that crawls, filters, and retrieves precise API endpoints to construct the prompt context.
First, we crawled the official COM API documentation for SolidWorks, Inventor, and AutoCAD. To minimize database noise, we filtered the APIs to retain those directly relevant to our evaluated tasks. For AutoCAD, we restricted the crawl to drawing and file-related collections. The statistics of the constructed API database are detailed in Table~\ref{tab:api_crawl_stats}, the retrieved API breakdowns are listed in Table~\ref{tab:task_retrieval_stats}, the standardized JSON schemas used to represent the APIs for each software are illustrated in Table~\ref{tab:api_json_format}.
Next, we prompt GPT-4o~\cite{ref:gpt-4o} to extract a set of technical sub-queries from the natural language instruction. These sub-queries are then embedded using OpenAI's text-embedding-3-large model and searched against our vectorized API database to retrieve a preliminary pool of candidate APIs.
Finally, we employ GPT-4o as a reranker. The reranker evaluates the candidate pool against the original sub-queries and selects the top 5 most relevant APIs per query. These finalized APIs—complete with their descriptions, parameters, and return types—are then injected into the prompt of the target LLM. The quantitative retrieval performance across different task categories is summarized in Table~\ref{tab:task_retrieval_stats}.

\begin{table*}[t]
\centering
\small
\textbf{AutoCAD}
\begin{verbatim}
{
    "type": "method",
    "parent_Object": "Document",
    "name": "SaveAs",
    "description": "Saves the document to a specified file.",
    "parameters": [],
    "return_value": "No return value."
}
\end{verbatim}
\vspace{0.5em}

\textbf{SolidWorks}
\begin{verbatim}
{
    "type": "method",
    "owner_interface": "ISldWorks",
    "name": "NewDocument",
    "description": "Creates a new document based on the specified template.",
    "parameters": [
        {
            "name": "TemplateName",
            "description": "Fully qualified path and name of the template..."
        },
        {
            "name": "PaperSize",
            "description": "Size of paper as defined in swDwgPaperSizes_e"
        }
    ],
    "return_value": "Newly created document or NULL if the operation fails"
}
\end{verbatim}
\vspace{0.5em}

\textbf{Inventor}
\begin{verbatim}
{
    "type": "method",
    "parent_Object": "FileManager",
    "name": "GetTemplateFile",
    "description": "Method that specifies the template to use when creating a file.",
    "parameters": [
        {
            "Name": "DocumentType",
            "Type": "DocumentTypeEnum",
            "Description": "Input constant that specifies the type of to create."
        },
        {
            "Name": "SystemOfMeasure",
            "Type": "SystemOfMeasureEnum",
            "Description": "Input constant that specifies the system of measure..."
        }
    ]
}
\end{verbatim}
\caption{Examples of the standardized JSON schemas used to represent crawled COM APIs in the RAG database. Differences in dictionary keys (e.g., \texttt{owner\_interface} vs. \texttt{parent\_Object}) naturally reflect the underlying COM architectures of the respective software.}
\label{tab:api_json_format}
\end{table*}

\section{Prompts}
\label{sec:prompts}
To evaluate the model's capabilities across different scenarios, we constructed three distinct system prompts: Baseline, Few-Shot, and Retrieval-Augmented Generation (RAG). All prompts share a foundational structure that defines the persona, execution rules, and input/output formats, but they differ in the provided context. The prompts used for our training and evaluation are shown in Table~\ref{tab:system_prompts}.

\begin{table*}[t]
\centering
\small
\begin{tabularx}{\textwidth}{@{}X@{}}
\toprule
\textbf{Baseline Prompt} \\
\midrule
\textbf{Role:} You are an expert CAD engineer and Python programmer specialized in COM interface automation. Your goal is to write executable Python codes that call the COM APIs (\texttt{win32com.client}) to programmatically drive CAD software (AutoCAD, Inventor, Solidworks) and complete the given CAD task. \newline
\vspace{-0.5em} \newline
\textbf{Rules:} \newline
- Think before coding. Generate deterministic, non-interactive codes. \newline
- Always output the complete, fully executable Python script. Do NOT provide partial snippets. \newline
- Use only \texttt{win32com.client}, \texttt{pythoncom}, and Python standard libraries. \newline
\vspace{-0.5em} \newline
\textbf{Inputs:} 1) Task instruction. 2) Terminal output from the last run. 3) Current screenshot after the last run. \newline
\vspace{-0.5em} \newline
\textbf{Outputs:} \newline
1) Brief reasoning inside \texttt{\textless thinking\textgreater...\textless/thinking\textgreater}. \newline
2) A high-level decision wrapped as: \texttt{```decision CODE (or DONE/FAIL) ```} \newline
3) If and only if the decision is CODE, output a single \texttt{```python ... ```} block. \\
\midrule
\textbf{Few-Shot Prompt (Appended to Baseline)} \\
\midrule
\textbf{Example: 3D Modeling in Solidworks} \newline
\textbf{Task Instruction:} Model this part in Solidworks: To construct the first part of the cylinder... \textit{[Detailed dimensions and constraints omitted for brevity]} ...export the model as an STL and STEP file. \newline
\vspace{-0.5em} \newline
\textbf{Output:} \newline
\texttt{\textless thinking\textgreater} \newline
Create a new part, select the Right Plane... \textit{[Reasoning Omitted]} \newline
\texttt{\textless/thinking\textgreater} \newline
\texttt{```decision} \newline
\texttt{CODE} \newline
\texttt{```} \newline
\texttt{```python} \newline
\texttt{import win32com.client} \newline
\texttt{\# ... [Full Code Omitted] ...} \newline
\texttt{```} \\
\midrule
\textbf{RAG Prompt (Appended to Baseline)} \\
\midrule
\textbf{External Knowledge Context:} \newline
Here are some COM APIs that might be useful for completing this task. \newline
\texttt{[} \newline
\texttt{\ \ \{} \newline
\texttt{\ \ \ \ "type": "method",} \newline
\texttt{\ \ \ \ "owner\_interface": "ISldWorks",} \newline
\texttt{\ \ \ \ "name": "NewDocument",} \newline
\texttt{\ \ \ \ \# ... [API details and other APIs Omitted] ...} \newline
\texttt{\ \ \}} \newline
\texttt{]} \\
\bottomrule
\end{tabularx}
\caption{System prompts used in our evaluation across Baseline, Few-Shot, and RAG settings. Unnecessary JSON arguments and code details are omitted for brevity.}
\label{tab:system_prompts}
\end{table*}

\end{document}